\documentclass[apj]{emulateapj}

    \usepackage[flushleft]{threeparttable}

\def\xmm{{{\it XMM-Newton}}}
\def\chandra{{\it Chandra}}
\def\nustar{{\it NuSTAR}}

\newcommand{\cgs}{ ${\rm erg~cm}^{-2}~{\rm s}^{-1}$} 
\newcommand{\lum}{\rm erg~s$^{-1}$}
\def\gtrsim{\mathrel{\hbox{\rlap{\hbox{\lower4pt\hbox{$\sim$}}}\hbox{$>$}}}}

\def\lum{{\rm {erg~s$^{-1}$}}}
\usepackage{txfonts}
\usepackage{natbib}

\bibpunct{(}{)}{;}{a}{}{,}
\slugcomment{Version: February 24, 2015}

\shorttitle{The \nustar\ COSMOS survey}
\shortauthors{Civano et al.}

\begin{document}


\title{The \nustar\ extragalactic surveys: overview and catalog from the COSMOS field}

\author{F. Civano\altaffilmark{1,2,3}, R. C. Hickox\altaffilmark{2}, S. Puccetti\altaffilmark{4}, A. Comastri\altaffilmark{5}, J. R., Mullaney\altaffilmark{6}, L. Zappacosta\altaffilmark{7}, S. M. LaMassa\altaffilmark{1}, J. Aird\altaffilmark{8,9}, D. M. Alexander\altaffilmark{8}, D. R. Ballantyne\altaffilmark{10}, F. E. Bauer\altaffilmark{11,12,13}, W. N. Brandt\altaffilmark{14,15,16},  S. E. Boggs \altaffilmark{17}, F. E. Christensen \altaffilmark{18}, W. W. Craig \altaffilmark{19,17}, A. Del-Moro\altaffilmark{8}, M. Elvis\altaffilmark{3}, K. Forster\altaffilmark{20}, P. Gandhi\altaffilmark{8,21}, B. W. Grefenstette\altaffilmark{20},  C. J. Hailey \altaffilmark{22}, F. A. Harrison\altaffilmark{20}, G. B. Lansbury\altaffilmark{8}, B. Luo\altaffilmark{11,12}, K. Madsen\altaffilmark{20}, C. Saez\altaffilmark{11}, D. Stern\altaffilmark{23}, E. Treister\altaffilmark{24}, M. C. Urry\altaffilmark{1}, D. R. Wik\altaffilmark{25}, W. Zhang \altaffilmark{25} } 

\altaffiltext{1}{Yale Center for Astronomy and Astrophysics, 260 Whitney Avenue, New Haven, CT 06520, USA}
\altaffiltext{2}{Department of Physics and Astronomy, Dartmouth College, 6127 Wilder Laboratory, Hanover, NH 03755, USA}
\altaffiltext{3}{Harvard-Smithsonian Center for Astrophysics, 60 Garden Street, Cambridge, MA 02138, USA}
\altaffiltext{4}{ASDC--ASI, Via del Politecnico, 00133 Roma, Italy}
\altaffiltext{5}{INAF--Osservatorio Astronomico di Bologna, via Ranzani 1, 40127 Bologna, Italy}
\altaffiltext{6}{The Department of Physics and Astronomy, The University of Sheffield, Hounsfield Road, Sheffield S3 7RH, U.K}
\altaffiltext{7}{INAF--Osservatorio Astronomico di Roma, via Frascati  33, 00040 Monte Porzio Catone (RM), Italy}
\altaffiltext{8}{Department of Physics, Durham University, South Road, Durham DH1 3LE, UK}
\altaffiltext{9}{Institute of Astronomy, University of Cambridge, Madingley Road, Cambridge CB3 0HA}
\altaffiltext{10}{Center for Relativistic Astrophysics, School of Physics, Georgia Institute of Technology, Atlanta, GA 30332, USA}
\altaffiltext{11}{Instituto de Astrof\'{\i}sica, Facultad de F\'{i}sica, Pontificia Universidad Cat\'{o}lica de Chile, 306, Santiago 22, Chile} 
\altaffiltext{12}{Millennium Institute of Astrophysics, Vicu\~{n}a Mackenna 4860, 7820436 Macul, Santiago, Chile} 
\altaffiltext{13}{Space Science Institute, 4750 Walnut Street, Suite 205, Boulder, Colorado 80301}
\altaffiltext{14}{Department of Astronomy and Astrophysics, The Pennsylvania State University, 525 Davey Lab, University Park, PA 16802, USA}
\altaffiltext{15}{Institute for Gravitation and the Cosmos, The Pennsylvania State University, 525 Davey Lab, University Park, PA 16802, USA}
\altaffiltext{16}{Department of Physics, The Pennsylvania State University, 525 Davey Lab, University Park, PA 16802, USA}
\altaffiltext{17}{Space Sciences Laboratory, University of California, Berkeley, CA 94720}
\altaffiltext{18}{DTU Space -- National Space Institute, Technical University of Denmark, Elektrovej 327, 2800 Lyngby, Denmark}
\altaffiltext{19}{Lawrence Livermore National Laboratory, Livermore, CA 945503}
\altaffiltext{20}{Cahill Center for Astronomy and Astrophysics, California Institute of Technology, 1216 E. California Blvd, Pasadena, CA, 91125 USA}
\altaffiltext{21}{School of Physics \& Astronomy, University of Southampton, Highfield, Southampton SO17 1BJ, UK}
\altaffiltext{22}{Columbia Astrophysics Laboratory, Columbia University, New York, NY 10027}
\altaffiltext{23}{Jet Propulsion Laboratory, California Institute of Technology, 4800 Oak Grove Drive, Pasadena, CA 91109, USA}
\altaffiltext{24}{Universidad de Concepci\'{o}n, Departamento de Astronom\'{\i}a, Casilla 160-C, Concepci\'{o}n, Chile}
\altaffiltext{25}{NASA Goddard Space Flight Center, Greenbelt, MD 20771}


\begin{abstract} 
To provide the census of the sources contributing to the X-ray background peak above 10 keV, \nustar\ is performing extragalactic surveys using a three-tier ``wedding cake'' approach. We present the \nustar\ survey of the COSMOS field, the medium sensitivity and medium area tier, covering 1.7 deg$^2$ and overlapping with both \chandra\ and \xmm\ data. This survey consists of  121 observations for a total exposure of $\sim$3 Ms. 
To fully exploit these data, we developed a new detection strategy, carefully tested through extensive simulations. The survey sensitivity at 20\% completeness is 5.9, 2.9 and 6.4 $\times$ 10$^{-14}$ \cgs\ in the 3-24 keV, 3-8 keV and 8-24 keV bands, respectively. By combining detections in 3 bands, we have a sample of 91 \nustar\ sources with 10$^{42}$-10$^{45.5}$ \lum\ luminosities and redshift $z$=0.04-2.5. Thirty two sources are detected in the 8-24 keV band with fluxes $\sim$100 times fainter than sources detected by {\it Swift}-BAT. Of the 91 detections, all but four are associated with a \chandra\ and/or \xmm\ point-like counterpart. One source is associated with an extended lower energy X-ray source. We present the X-ray (hardness ratio and luminosity) and optical-to-X-ray properties. The observed fraction of candidate Compton-thick AGN measured from the hardness ratio is between 13\%-20\%. We discuss the spectral properties of \nustar\ J100259+0220.6 (ID~330) at $z$=0.044, with the highest hardness ratio in the entire sample. The measured column density exceeds 10$^{24}$~cm$^{-2}$, implying the source is Compton-thick. This source was not previously recognized as such without the $>$10 keV data. 
\end{abstract}

\section{Introduction}

 
For more than 30 years, X-ray surveys have provided a unique and powerful tool to find and study accreting supermassive black holes (SMBHs) in the distant Universe (Fabian \&\ Barcons 1992, Brandt \&\ Hasinger 2005, Alexander \& Hickox 2012, Brandt \&\ Alexander 2015). In the past decade alone, dozens of surveys with \xmm\ and \chandra\ have covered a wide range in area and X--ray flux, corresponding to a similarly wide range in luminosity and redshift. The luminosity function of Active Galactic Nuclei (AGN) has thus been sampled over three decades or more in X--ray luminosity and up to redshifts $z$=5 (Ueda et al. 2003, La Franca et al. 2005, Hasinger 2008, Brusa et al. 2009, Civano et al. 2011, Vito et al. 2013, Ueda et al. 2014, Kalfountzou et al. 2014), defining the evolution of  unobscured ($N_{\rm H} < 10^{22}$~cm$^{-2}$) and obscured ($ N_{\rm H} >10^{22}$~cm$^{-2}$) sources and reaching fainter luminosities than optical surveys.
 
However, these surveys are biased against the discovery of heavily obscured accreting SMBHs enshrouded by gas with column densities greater than the inverse of the Thompson scattering cross-section, called Compton-thick AGN (hereafter CT; $N_{\rm H} > 10^{24}$~cm$^{-2}$), at $z<$1.5, as \chandra\ and \xmm\ are most sensitive in the 0.5--8 keV energy range, where the emitted X-rays can be absorbed by high column densities of intervening matter. Non-focusing hard X-ray satellites have performed population studies, such as those obtained using data from {\it RXTE} (Sazonov \& Revnivtsev 2004), {\it INTEGRAL}-IBIS (Beckmann et al. 2006) and {\it Swift}-BAT (Tueller et al. 2008, Burlon et al. 2011). However, other than highly beamed blazars, sources found in the previous studies were generally of moderate luminosities and restricted to the nearby ($z < 0.2$) universe.
The importance of this population is recognized but the fraction of CT AGN is still currently highly uncertain as is their contribution to the X-ray background emission at its 20-40 keV peak. Three unequivocal signatures of heavy obscuration in the X-rays are: (i) the presence of absorption at low X-ray energies (E$<$10~keV), (ii) high equivalent width iron lines and edges (E=6--7~keV), and/or (iii) a Compton-scattered reprocessing hump in the E$>10$~keV X-ray spectrum. Therefore, observations at energies above 10~keV are essential to fully understand the intrinsic  emission of the most heavily obscured AGN. The relative strength of the various components and the effect of Compton scattering on the absorbing column density are best studied at E $>$ 10 keV. 
 
The {\it Nuclear Spectroscopic Telescope Array} (\nustar, Harrison et al. 2013) with its novel focusing capabilities at $>$ 10 keV and angular resolution of 18$^{\prime\prime}$ (full width half maximum) provides a unique opportunity to detect and study obscured and CT AGN out to moderate redshifts ($z\sim1$). \nustar\ probes down to a limiting flux more than two orders of magnitude fainter than possible with previous (non-focussing) hard X-ray surveys by {\it Swift}-BAT and {\it INTEGRAL}-IBIS (Krivonos et al. 2010, Burlon et al. 2011, Baumgartner et al. 2013).
 
A ``wedding cake'' strategy for the \nustar\ surveys consisting of different areas observed and different depths was designed to unveil this heavily obscured population and to determine the distribution of absorbing column densities in the AGN population as a whole (Harrison et al. in prep.). Three major components of this wedding cake include: deep (3$\times$10$^{-14}$ \cgs\ in the 3-24 keV band) surveys over the Extended \chandra\ Deep Field South (ECDFS; Mullaney et al.  2015; $\sim$0.33 deg$^2$) and the region of deepest \chandra\ exposure (Goulding et al. 2012) in the Groth Strip (Aird et al. in prep.; $\sim$0.25 deg$^2$); a medium depth survey over the Cosmic Evolution Survey field (COSMOS; Scoville et al. 2007) reported here; and a large area survey (currently covering $\sim$6 deg$^2$) including all serendipitous sources discovered in non-survey fields (Alexander et al. 2013, Lansbury et al. in prep.). 
In the ECDFS survey, \nustar\ was able to classify source J$033202-274650$ (Del Moro et al. 2014) as highly obscured and place better constraints on the obscuring column and spectral shape. This object was not clearly identified as having a high, but Compton-thin, column density (5.6$\times$10$^{23}$ cm$^{-2}$) using only \chandra\ or \xmm\ data. Also, \nustar\ observations of SDSS selected candidate CT quasars robustly measured their obscuration level (Lansbury et al. 2014, submitted).

Here we present the $\sim$1.7 deg$^2$ \nustar\ COSMOS survey, which overlaps the region covered by \xmm\ and \chandra\ at lower energies ({\it XMM}-COSMOS: Hasinger et al. 2007, Brusa et al. 2010; C-COSMOS: Elvis et al. 2009, Civano et al. 2012; \chandra\ COSMOS Legacy, Civano et al. in prep., Marchesi et al. in prep.). In this paper, we describe the \nustar\ observations (\S~2), data processing (\S~3), extensive simulations carried to assess the reliability of detected sources and the sensitivity of the survey (\S~4), data analysis and properties of the detected sources (\S~5). In section \S~6, we present in detail the spectral analysis of source ID 330 at $z$=0.044, with the most extreme hardness ratio in the whole sample. Source ID 330 is a new CT AGN in the local Universe. In the Appendix, we present the point source catalog. 

We assume a cosmology with H$_0$ = 71~km~s$^{-1}$~Mpc$^{-1}$, $\Omega_M$ = 0.3 and $\Omega_{\Lambda}$= 0.7. 

\section{Observations}

The \nustar\ survey of the COSMOS field consists of 121 overlapping fields with each tile (12$^{\prime}\times$12$^{\prime}$) shifted by half a field, forming an 11$\times$11 grid. This strategy of half shifts, used also in C-COSMOS and in the \chandra\ COSMOS Legacy survey, results in a relatively uniform exposure over the observed area.
The \nustar\ COSMOS survey was performed during three different periods in 2013 and 2014: the first 34 observations were taken between the end of December 2012 and the end of January 2013, the next 30 observations were taken during the months of April and May 2013 and the last 57 observations were taken between December 2013 and February 2014. Observation details, including pointing coordinates, roll angles, observing dates and exposure times\footnote{The exposure times reported here are those already cleaned and corrected for high background events (see Section \ref{data_reduction}).} for both focal plane modules A (FPMA) and B (FPMB; Harrison et al. 2013) are reported in an electronic Table. 

The timing of the observations allowed the roll angle to be kept nearly constant, differing by a maximum of 30 deg (and flipped by 180 deg). The exposure time of each pointing ranges between 20 and 30 ksec, with three fields split into multiple observations due to satellite scheduling constraints. The total exposure allocated so far for the COSMOS field is $\sim$3.12 Ms. This tiling strategy produces a deeper inner area of $\sim$1.2 deg$^2$ with an average (vignetting corrected) exposure time of $\sim$50-60 ks for each module and an outer frame with half of the inner exposure covering $\sim$0.5 deg$^2$ (see Figure \ref{expo_area}).

\section{Data Processing}

\subsection{Data reduction}
\label{data_reduction}

We performed the \nustar\ data reduction using HEASoft v.6.15.1 and the \nustar\ Data Analysis Software 
NuSTARDAS v.1.3.1 with CALDB v.20131223 (Perri et al. 2013\footnote{http://heasarc.gsfc.nasa.gov/docs/nustar/analysis/nustar\_swguide.pdf}). 
We processed level 1 products by running {\it nupipeline}, a NuSTARDAS module which includes all the necessary data reduction steps to obtain a calibrated and cleaned event file ready for scientific analysis. A further bad pixel file was also used in the filtering to flag outer edge pixels with high noise.


Light curves were produced using {\it nuproducts} in the low--energy range (3.5-9.5 keV), with a 500 second interval binning. Twenty-one observations (fields 13, 20, 21, 55 , 56, 57, 58, 70, 83, 87, 88, 89, 91, 92, 96, 97, 110, 111, 116, 119, 120) were affected by abnormally high background radiation due to solar flares. Intervals with count rates $>$0.1 cnt~s$^{-1}$ were cleaned by reprocessing the twenty-one observations, applying good time interval files created using standard HEASoft tools. The resulting loss of exposure time corresponds to 2-9\% of the exposure in these twenty-one observations, for a total of 21 ksec. This is less than 1\% of the total \nustar\ COSMOS exposure.

\subsection{Exposure map production}

Exposure maps were created using {\it nuexpomap}, which computes the net exposure time for each sky pixel, for a given 
observation. In order to reduce the calculation time, we binned the maps using a bin size of 5 pixels. The exposure map accounts for bad and hot pixels, detector gaps, attitude variations and mast movements.  
Ideally, given the strong energy dependence of the effective area, exposure maps at each energy should be created and then summed together using weights based on the effective area at each energy. This is more complicated if the typical source spectrum is not constant with energy. This procedure is computationally expensive and it can be simplified by convolving the instrument response with a specific model for the incident spectrum.
The energies at which the exposure maps were created were computed by weighting the effective area with a power law spectrum with $\Gamma$=1.8 (see Section \ref{simul1} for spectral choice motivation). The mean, spectrally weighted energies are 5.42 keV for the 3-8 keV band, 13.02 keV for the 8-24 keV band and 9.88 keV for the 3-24 keV band. 

\begin{figure}
\centering
\includegraphics[width=0.45\textwidth, angle=0]{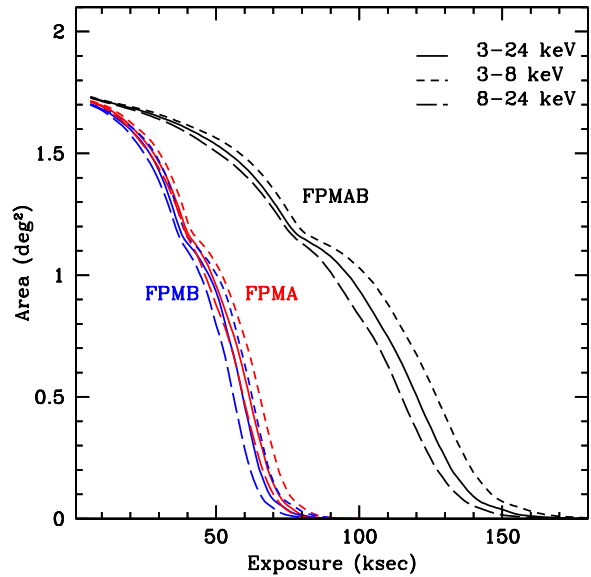}
\caption{\small Survey area coverage as a function of the effective, i.e. vignetting-corrected, exposure time for FPMA (red), FPMB (blue) and the sum of the two (black) in three energy bands: 3-24 keV (solid line), 3-8 keV (short dashed line) and 8-24 keV (long dashed line). }
\label{expo_area}
\end{figure}

The 121 vignetting corrected exposure maps were mosaicked using the HEASoft XIMAGE tool. The effective exposure time, corrected for vignetting, is plotted versus area in Figure~\ref{expo_area}.
A small difference of $<$5\% in exposure per area is seen between FPMA and FPMB, with FPMA being more sensitive.

\subsection{Background map production}
 \label{sec:back_map}
 
As described in detail in Wik et al. (2014), the \nustar\ observatory has several independent background components which vary
spatially across the field of view.
The background spectrum can be decomposed into five components of fixed spectral shape (excluding instrumental line strengths) with individual normalizations that are position dependent. Below 20 keV, the background is dominated by stray light from unblocked sky emission leaking
through the aperture stop. This component is by nature spatially non-uniform. 
Below 5 keV, there is a spatially uniform component across all detectors, due to emission from unresolved X-ray sources focused by the optics (denoted fCXB for focused Cosmic X-ray Background). 
A further low-energy background component is related to solar photons reflecting off the back of the aperture stop, producing 
a time-variable signal related to whether or not the instrument is in sunlight.
 Above 20 keV, the background is dominated by the detector emission lines produced by interactions between the spacecraft/detectors and the radiation
environment in orbit, as well as several fluorescence lines. 

We used {\it nuskybgd} (Wik et al. 2014) to produce accurate background maps for each observations taking into account all the components described above. Using this code, we extracted spectra (and response matrices) in four circular regions, 
covering each quadrant of the field of view (with radius of 2.8$^{\prime}$) and avoiding the gaps between the detectors for both FPMA and FPMB for each observation. The eight extracted spectra were jointly fitted with XSPEC v.12.8.1 (Arnaud 1996) to determine the normalization of all the above components in each observation. Because the fCXB component is more than 10 times fainter than the aperture background component, we first fit a fixed normalization to this component (using the nominal value from {\it HEAO-1} measured normalization, Boldt 1987) and then we let it vary once the other components were constrained. Given the overall small number of counts in the background spectra ($\sim$ 1000 counts), the Cash statistic (Cash 1979) was used for the spectral fitting. 
Background maps were then produced using the fitted normalizations.  

\begin{figure}
\centering
\includegraphics[width=0.45\textwidth, angle=0]{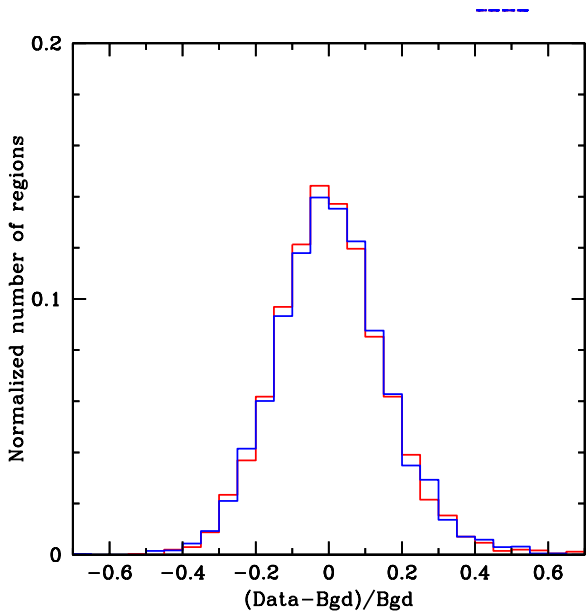}
\caption{\small Normalized distributions of relative differences between counts extracted from observed data and background maps in each tile for FPMA (red) and FPMB (blue).}
\label{backmaphisto}
\end{figure}

To compare the generated background maps and the background value measured in the data images, we extracted the counts from the background maps to compare with the counts extracted in the same regions from the observed data for both FPMA and FPMB. We covered each field with 64 regions of 45$^{\prime\prime}$ radius.
Figure~\ref{backmaphisto} presents the normalized distribution of relative differences between background and observed data counts extracted in each region in each field (red: FPMA; blue: FPMB). Given the absence of bright \nustar\ sources in the COSMOS field, we find that removing detected sources when computing the background maps does not significantly change the overall background distribution. From Gaussian fitting of the distributions of the difference between source and background shown in Figure~\ref{backmaphisto}, we find centroids at $(Data-Bgd)/Bgd$ =0.0023 and 0.0047 and standard deviations of 0.144 and 0.147 for FPMA and FPMB, respectively, showing a remarkable agreement between the generated background maps and the data.

\subsection{Astrometric Offsets}
\label{astrometry}

Any errors in the astrometric solution for the different exposures can introduce a loss of sensitivity due to the decrease of angular resolution when summing the observations into the mosaic. 
We therefore tested whether significant astrometric offsets 
are affecting our observations. 
The catalogs of detected \chandra\ sources in COSMOS (Elvis et al. 2009, Civano et al. 2015 submitted) was used as the reference for computing astrometric offsets.
However, given the lack of multiple bright sources in individual observations required to perform accurate alignments, we used a stacking technique applied to FPMA and FPMB separately. 
At first, \chandra\ sources in each individual \nustar\ observation were stacked (by removing neighbors and keeping brighter sources) but the resulting signal in most fields had insufficient signal-to-noise ratio ($<$1) to provide an accurate astrometric correction. We then stacked \chandra\ sources in contemporaneous and contiguous \nustar\ observations, i.e. during which the telescope did not move to observe another target between one COSMOS field and the next. This procedure assumes a stable alignment of the observatory. 
The astrometric offsets measured for stacked sources with high signal-to-noise ratio ($>$3) were in the range 1$^{\prime\prime}$ -7$^{\prime\prime}$, comparable to the \nustar\ pixel size and consistent with expected uncertainties (see also Section~\ref{rel}). Therefore we decided not to perform any astrometric corrections to our data. No significant offsets were found between FPMA and FPMB.

\subsection{Mosaic creation}
\begin{figure*}
\centering
\includegraphics[width=0.9\textwidth, angle=0]{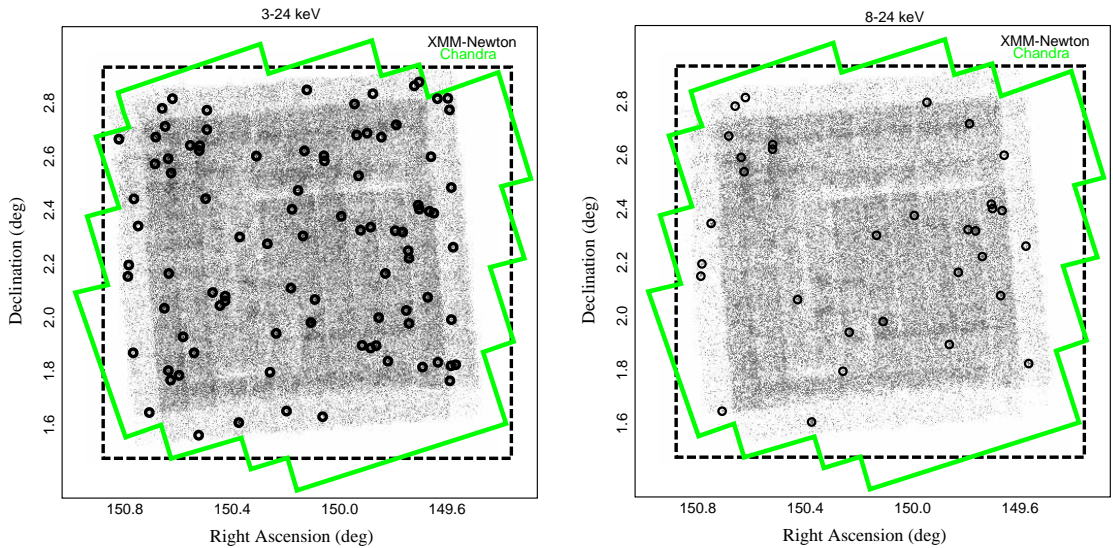}
\caption{\small Combined FPMA+B 3-24 keV (left) and 8-24 keV (right) band mosaics with the positions of the 91 detected sources on the left and the 8-24 keV detected sources on the right, as black circles. The \chandra\ and \xmm\ surveys area are marked as green solid and black dashed polygons, respectively.}
\label{mosaic}
\end{figure*}

The 121 observations were merged using the HEASoft tool XSELECT into three mosaics: FPMA, FMPB and the summed FPMA+B. Following Alexander et al. (2013), the FPMA, FMPB and FPMA+B event mosaics were filtered in energy using the CIAO (Fruscione et al. 2006) tool {\it dmcopy} into three bands, 3-8 keV, 8-24 keV and 3-24 keV.
The high energy limit of 24 keV for the analysis has been imposed by the presence of relatively strong instrumental lines at 25-35 keV, whose parametrization is uncertain which can lead to spurious high residuals during the modeling of the background. Source detection at energies above 35 keV is possible but is beyond the scope of our paper, and will be the subject of a future work.
In order to achieve the deepest sensitivity, we performed detection and analysis on the merged FPMA+B mosaic, as we are confident of the alignment of the two detectors (Section \ref{astrometry}). The 3-8 keV and 8-24 keV band mosaics are shown in Figure~\ref{mosaic} compared to the area covered by \chandra\ and \xmm\ on the same field. The \nustar\ COSMOS survey is fully covered by both lower energy X-ray telescopes.

\section{Simulations}
\label{simul}

Extensive simulations were performed in order to develop, test and optimize the source detection methodology. Moreover, the simulations were used 
to estimate the level of significance of each detected source, 
to determine the level of completeness of the source list as a function of source flux, the reliability of sources as a function of source significance, 
and the detected source position accuracy.

\subsection{Generation of simulated data}
\label{simul1}

To generate simulated maps, mock sources were assigned fluxes drawn randomly from the number counts estimated assuming the Treister et al. (2009) model in the 3-24, 3-8 and 8-24 keV bands, using an online number count Monte Carlo calculator\footnote{http://agn.astro-udec.cl/j\_agn/main.html}. Given that we are simulating the total population and not a sub sample of sources, we can assume that the number count model used in this work is consistent with other models available in the literature as Gilli et al. (2007), Akylas et al. (2012) and Ballantyne et al. (2011). 

The minimum flux for the input catalog was 5$\times$10$^{-15}$\cgs\ in the 3-24 keV band, which is a factor of $\sim$10 below the expected \nustar\ COSMOS limit. Hence, background fluctuations due to unresolved sources are included in the simulations. Fluxes for the 3-8 and 8-24 keV bands were computed using a power-law model with slope $\Gamma$=1.8, the typical value for AGN in this energy range (Burlon et al. 2011, Alexander et al. 2013), and Galactic column density $N_{\rm H}$=2.6$\times 10^{20}$ cm$^{-2}$ (Kalberla et al. 2005). The counts-to-flux conversion factors (CF) adopted here, computed using the response matrix and ancillary file available in the adopted CALDB, are CF= 4.59, 3.22 and 6.64 $\times$10$^{-11}$ ${\rm erg~cm}^{-2}~{\rm counts}^{-1}$ in the 3-24, 3-8 and 8-24 keV bands, respectively.

These sources were randomly added to a background map produced as described in Section~\ref{sec:back_map}, though without the fCXB component included. 
The point spread function (PSF) used to add sources to the background map is taken from the \nustar\ PSF map available in the CALDB. It changes as a function if the off-axis angle (and azimuthal angle), and is the sum of the PSFs for all the observations covering any certain position.
Then, to match the total number of counts in the observed \nustar\ mosaic, additional background counts (4-6\% of the total) were added by scaling a background map with only the fCXB component for which the normalization has been averaged over all the fields. These additional counts are due to the fact that we averaged the fCXB component normalization, to sources fainter than 5$\times$10$^{-15}$\cgs\ which are not included in the simulation, and also to different spectral shapes of the true source population. A Poisson realization of the final map (sources plus background) was made. With the above procedure, we simulated a set of 400 mosaics in three bands (3-24, 3-8 and 8-24 keV) for both FPMA and FPMB, which were then summed to create simulated FPMA+B mosaics. 


\subsection{Simulation source detection and photometry}
\label{sourcedetection}

In order to fully exploit the large area and depth of the \nustar\ COSMOS survey, a dedicated analysis procedure for detection and source reliability was developed and applied to the simulated dataset with the aim of validating it. The same procedure, described below, was then applied to the real data (see Section \ref{realdata}) as well as to the ECDFS data in Mullaney et al. (2015).

Following Mullaney et al. (2015), we used {\it SExtractor} (Bertin \&\ Arnout 1996) to obtain a large catalog of potential sources. 
The source detection was performed on false probability maps generated by convolving the data mosaic (either simulated or observed) and the corresponding background map (the mosaic where the normalization of the fCXB component has been averaged over all the fields) using a circular top-hat function with two smoothing radii (10$^{\prime\prime}$, 20$^{\prime\prime}$) to detect sources with different sizes, i.e. to take into account overlapping point sources across the mosaic. 
To convert the convolved maps into Poisson probability maps, we used the incomplete gamma function, {\it igamma} (available in IDL), so that P$_{random} =igamma(Sci,Bgd)$, 
where Sci and Bgd are the smoothed science and background mosaics. These probability maps give the likelihood that the signal at each position in the mosaic is due to random background fluctuations. We computed the logarithm of these maps and inverted them so that significant fluctuations are positive. 
These maps were then input to {\it SExtractor} using a detection significance of 10$^{-4.5}$, set to avoid the loss of any real but faint detection.  

We performed aperture photometry at the positions obtained by {\it SExtractor} for each detected source. Total source counts were extracted from the data mosaic and background counts were extracted from the background mosaic in the 3-24, 3-8 and 8-24 keV bands, using a circular aperture of 20$^{\prime\prime}$ radius. With total and background counts, we computed the maximum likelihood (DET\_ML) for each source (see Puccetti et al. 2009; Cappelluti et al. 2009; LaMassa et al. 2013 for similar approaches). The DET\_ML  is related to the Poisson probability that a source candidate is a random fluctuation of the background (P$_{random}$): DET\_ML=$-ln$~P$_{random}$. Sources with low values of DET\_ML, and correspondingly high values of P$_{random}$, are likely to be background fluctuations. 
Detection and photometry were both computed in the three different bands separately. 

The sources detected in each probability map were merged into a single list, and duplicate sources were removed using a matching radius of 30$^{\prime\prime}$, i.e., if there are two sources with a separation smaller than 30$^{\prime\prime}$ only the one with the higher DET\_ML  is kept in the catalog. Given the size of the point spread function, a 30$^{\prime\prime}$ matching radius should not produce a large number of false matches ($<$few \%). The mean number of sources detected in each of the probability maps (in each band) and the final number of sources after cleaning the list of multiple detections of the same source are reported in Table~\ref{detections}.

\begin{table}[]
\small
\centering
\caption{Mean number of detected sources with {\it SExtractor} in simulated smoothed (10$^{\prime\prime}$ and 20$^{\prime\prime}$ radii) maps (line 1 to 3) , of detected sources matched to an input catalog source within 30$^{\prime\prime}$ (line 4), of detected sources with DET\_ML$>$ threshold (line 5).}
\begin{tabular}{l c c c c c c}
\hline
\hline
 & 3-24 keV &3-8 keV & 8-24 keV\\
\hline

10$^{\prime\prime}$ smoothed maps & 259 &219  & 152  \\
20$^{\prime\prime}$ smoothed maps & 227 & 193 &  123\\
Combined              & 269  & 230 & 173   \\
Matched to input  & 179 (66\%) & 167 (72\%) & 103 (60\%) \\
DET\_ML$>$ DET\_ML(99\%) &77 & 62 & 27\\ 
\hline
\hline

\end{tabular}
\label{detections}
\end{table}

We then used the procedure described by Mullaney et al. (2015; their section 2.3.2) to deblend the counts of sources which have been possibly contaminated by objects at separations of 90$^{\prime\prime}$ or lower. Deblended source and background counts were used to compute new DET\_ML values for each source.

We thus obtained for each band a catalog of detected sources which was then matched to the list of input mock sources using a positional cross-correlation, with a maximum separation of 30$^{\prime\prime}$. We report the mean number of detected sources matched to an input catalog source for each band in the fourth line of Table~\ref{detections}. The fraction of matched sources with respect to detected sources is $\sim$66\%, 72\% and 60\% in the 3-24 keV, 3-8 keV and 8-24 keV bands, respectively. 

\begin{figure}
\centering
\includegraphics[width=0.35\textwidth, angle=0]{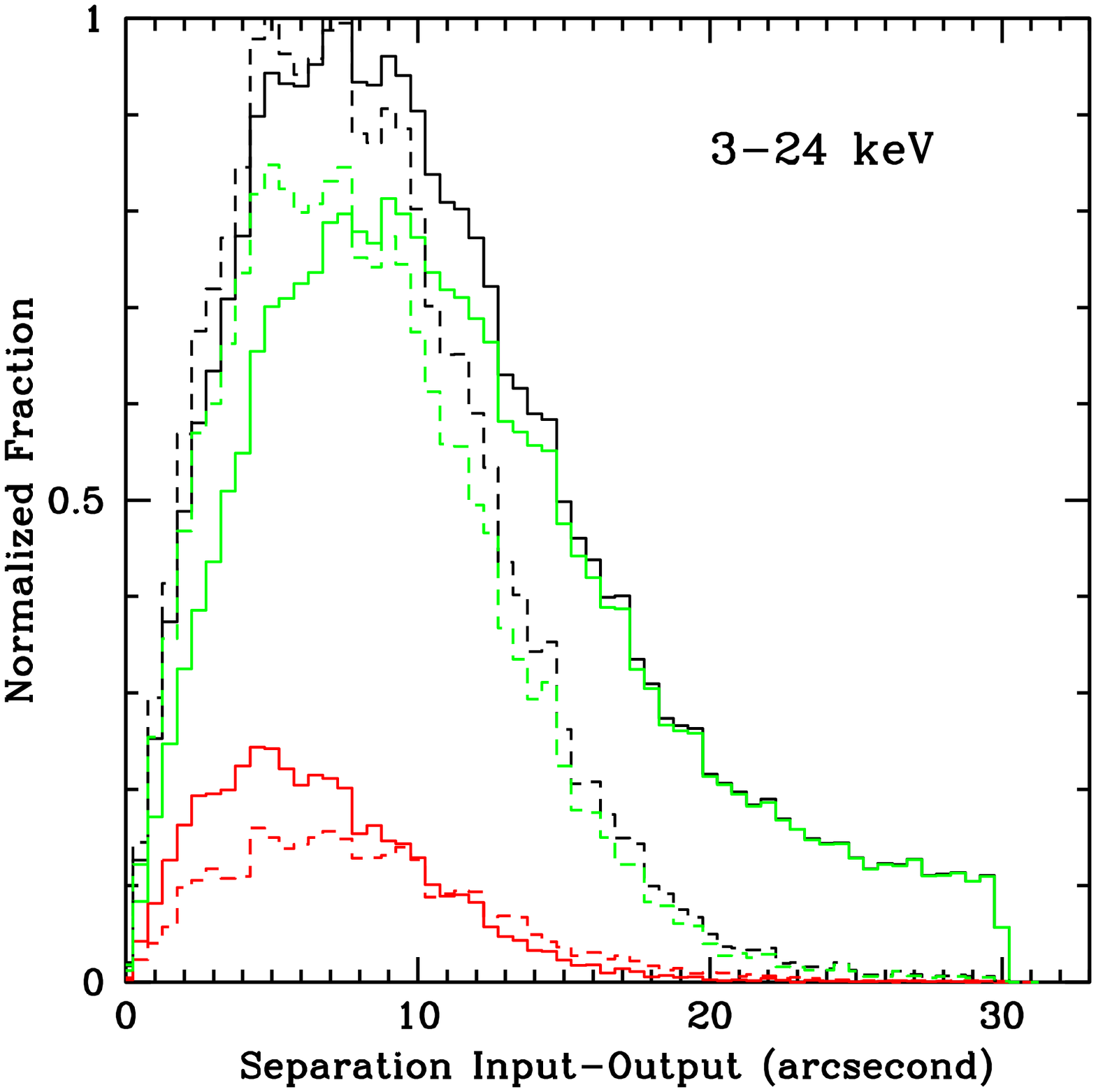}
\includegraphics[width=0.35\textwidth, angle=0]{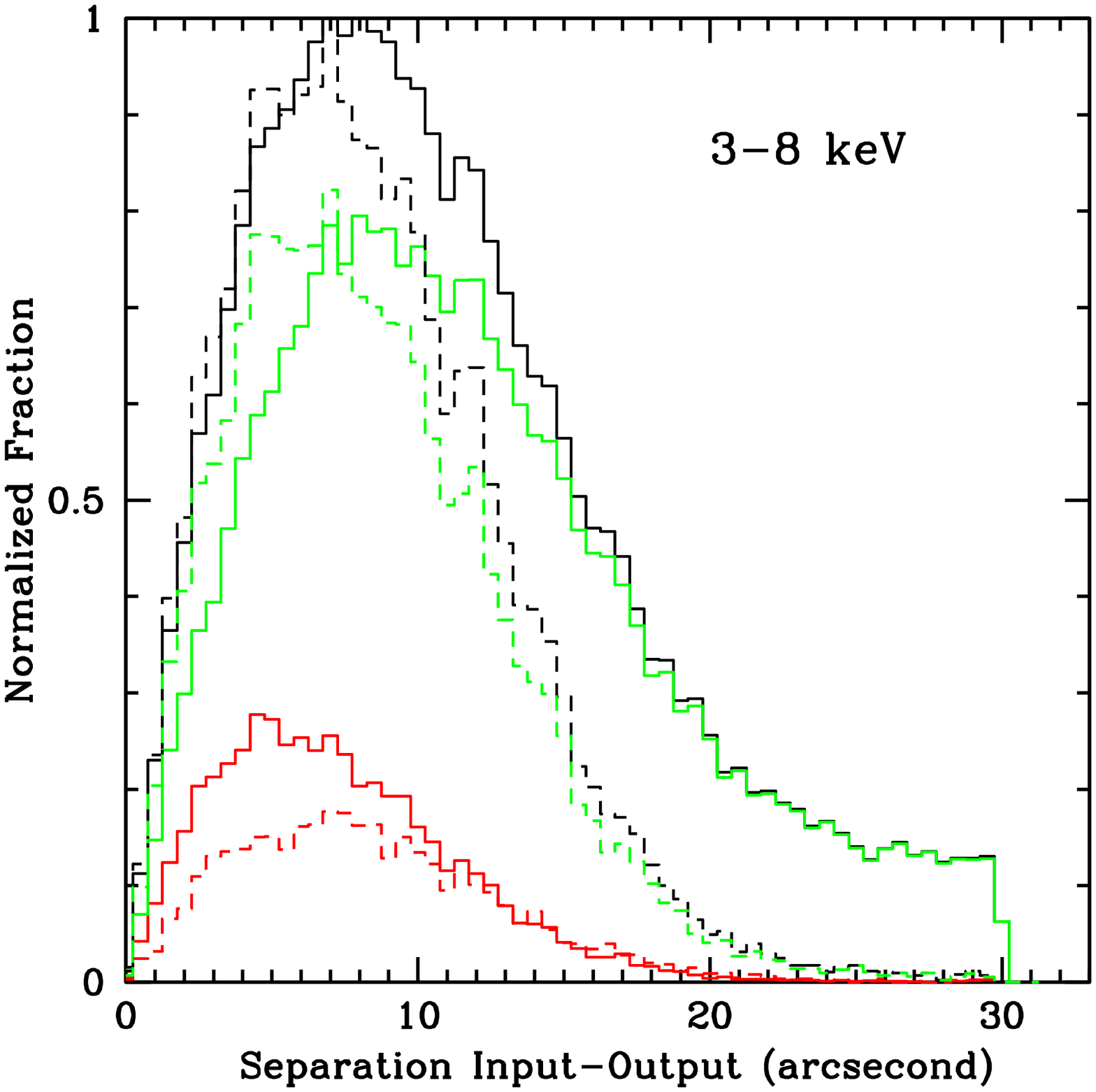}
\includegraphics[width=0.35\textwidth, angle=0]{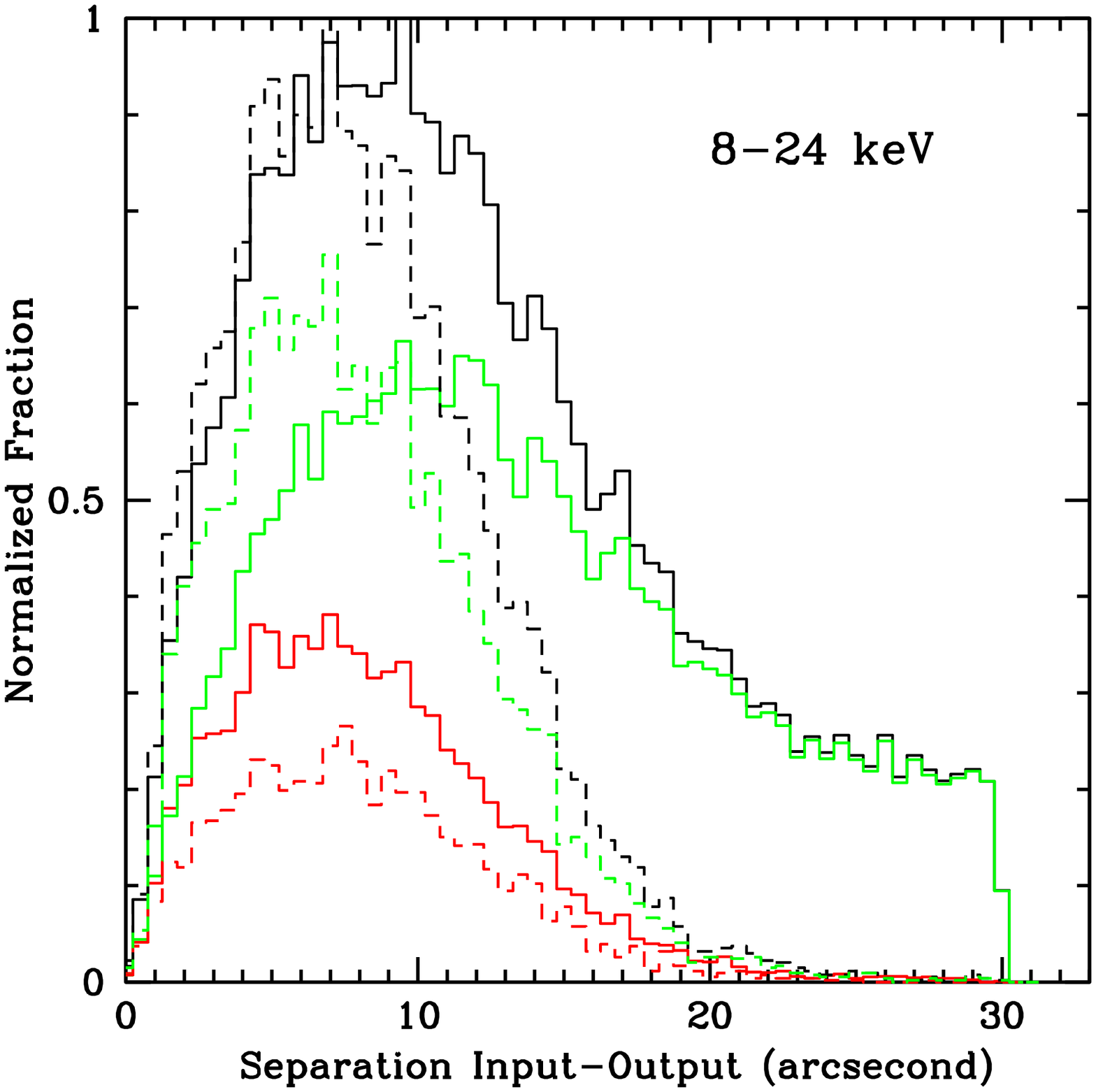}
\caption{\small Normalized distributions of the separations, in the simulations, between the detected positions obtained with {\it SExtractor} and the input positions in the 3-24 keV (top), 3-8 keV (middle) and 8-24 keV (bottom) bands. Solid and dashed lines refer to the whole sample and only to sources above the 99\% reliability DET\_ML thresholds in each band. The black lines represent the total distribution. The red lines represent the separations for bright sources with 3-24 keV flux $\geq$10$^{-13}$ \cgs\ and the green lines represent sources with fluxes $<$10$^{-13}$ \cgs.}
\label{distances}
\end{figure}

In Figure~\ref{distances}, we plot the distribution of the separation between input source positions and detected source positions for the three bands. These distributions are flux dependent: the distribution of bright sources (green lines, $\geq$10$^{-13}$ \cgs) peaks at $\sim$5$^{\prime\prime}$, while the one of fainter ($<$10$^{-13}$ \cgs) sources peaks at 8$^{\prime\prime}$. 
In the 3-24 and 3-8 keV bands, $\sim$55\% of the matches are within 10$^{\prime\prime}$ and $\sim$90\% within 20$^{\prime\prime}$. These numbers are slightly lower in the 8-24 keV band, where $\sim$45\% of the matches are within 10$^{\prime\prime}$ and $\sim$85\% within 20$^{\prime\prime}$, because of the lower number of counts in this band, however the fraction of sources within 20$^{\prime\prime}$ is still very high. The 30$^{\prime\prime}$ matching radius was chosen a posteriori to avoid losing the tail of sources at large separations.

\subsection{Reliability, completeness and survey sensitivity}
\label{rel}

The threshold for source detection must be set by balancing reliability and completeness. 
Reliability, which is an indicator of the number of spurious sources in the sample, is computed using the ratio between the number of detected sources matched to an input source and the total number of detected sources. Completeness is instead the ratio between the number of detected real sources and the number of input simulated sources. 
The analysis of the simulations allows us to choose a threshold in source significance, or DET\_ML, to maximize the reliability of the sources in the sample, while simultaneously maximizing completeness. A lower DET\_ML threshold gives higher completeness at the cost of lower reliability. 

\begin{figure}
\centering
\includegraphics[width=0.45\textwidth, angle=0]{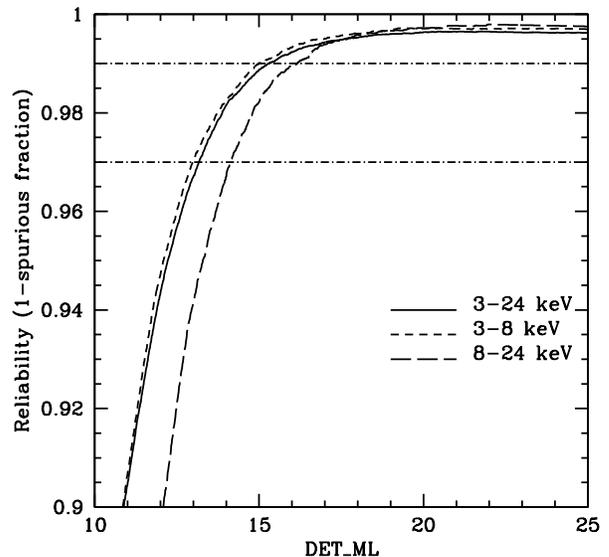}
\caption{\small Reliability as a function of DET\_ML from FPMA+B simulation analysis. Solid line: 3-24 keV band; short dashed line: 3-8 keV band; long dashed line: 8-24 keV band. The horizontal dot dashed lines represent the 99\% and 97\% reliability thresholds. }
\label{reliability}
\end{figure}

\begin{figure}
\centering
\includegraphics[width=0.45\textwidth, angle=0]{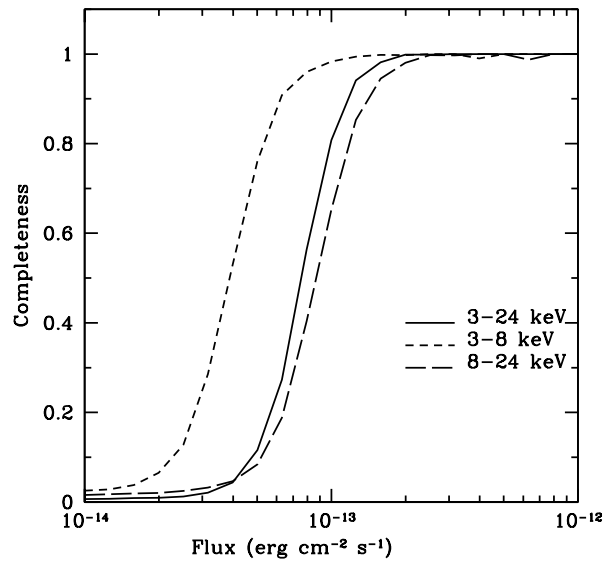}
\caption{\small Completeness as a function of X-ray flux at 99\% reliability DET\_ML threshold. Solid line: 3-24 keV band; short dashed line: 3-8 keV band; long dashed line: 8-24 keV band.}
\label{completeness}
\end{figure}

Figure~\ref{reliability} shows the cumulative distribution of reliability as function of DET\_ML in three bands for the FPMA+B simulations.
The horizontal lines represent 97\% and 99\% reliability (e.g., 3\% and 1\% spurious sources) . 
For what follows, we use the significance level corresponding to 99\% reliability: DET\_ML(99\%)=15.27, 14.99 and 16.17 for the 3-24 and 3-8 keV and 8-24 keV bands, respectively, corresponding to Poisson probabilities of $log_{10}$P$_{random}$=-6.63,-6.51,-7. 
Figure~\ref{completeness} shows the completeness in the three bands as a function of  X-ray flux at the DET\_ML threshold corresponding to 99\% reliability. 
Table~\ref{limit_flux} gives the flux limits corresponding to four completeness fractions in the three bands.
The mean numbers of detections above the 99\% reliability DET\_ML threshold found in the simulations are listed in Table~\ref{detections} (last row). The fraction of detected sources 
above this threshold is between 35-40\% in all bands.

\begin{figure}
\centering
\includegraphics[width=0.45\textwidth, angle=0]{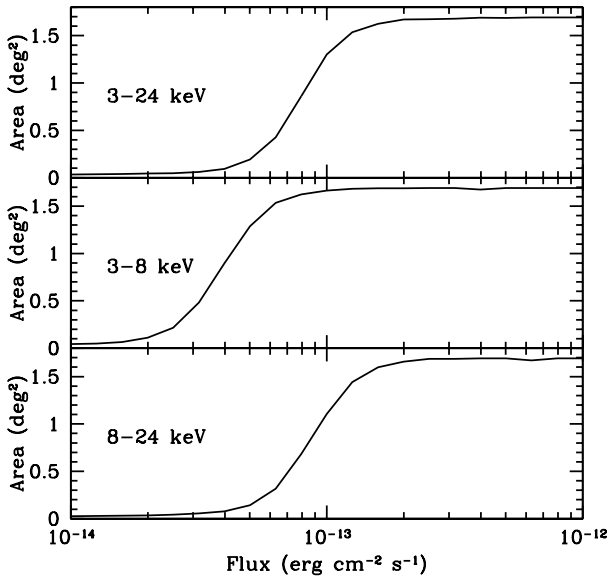}
\caption{\small Sky coverage at 99\% reliability DET\_ML level. Top: 3-24 keV band; middle: 3-8 keV band; bottom: 8-24 keV band.}
\label{sensitivity}
\end{figure}

The ``sky-coverage'' is the integral of the survey area covered down to a given flux limit. If at the chosen detection threshold the completeness is sufficiently high (with reliability also high), the number of detected sources should correspond to the number of input sources with DET\_ML higher than the threshold value. In this case, the curves in Figure~\ref{completeness} represent normalized
sky-coverages. The survey ``sky-coverage'' in the three bands is plotted in Figure~\ref{sensitivity}. 

In Figure~\ref{distances} (dashed lines), we plot the distribution of the separation between input and detected source positions restricted to the matches above the DET\_ML threshold, for the three bands. The fraction of matches within 10$^{\prime\prime}$ and 20$^{\prime\prime}$ is significantly improved when considering only sources with DET\_ML above the 99\% reliability threshold, increasing for all the bands to 68\% and 98\%, respectively. Comparing to the distribution of the whole sample (solid line) it is possible to see that the tail at large separations is made by sources at low DET\_ML values. By performing a two-dimensional Gaussian fitting of the distributions, we find a consistent width in all bands, and at both bright and faint fluxes of 6.6$^{\prime\prime}$, which can therefore be associated with the positional uncertainty of the detections. This value is consistent with what is expected for the source positional uncertainties according to the minimum resolvable separation Rayleigh criterion, assuming instead of the first diffraction minimum of a point source, the size corresponding to 20\% of the encircled energy fraction ($\sim$10$^{\prime\prime}$).

\begin{table}[]
\footnotesize
\centering
\caption{Completeness as function of flux (for 99\% reliability catalog)}
\begin{tabular}{l c c c c c c}
\hline
\hline
Completeness & F(3-24 keV) & F(3-8 keV)  & F(8-24 keV) \\
 &  \cgs\ &  \cgs\ &  \cgs\ \\
90\% & 1.1 $\times$ 10$^{-13}$ & 6.1 $\times$ 10$^{-14}$ & 1.3 $\times$ 10$^{-13}$ \\ 
80\% & 1.0 $\times$ 10$^{-13}$ & 5.3 $\times$ 10$^{-14}$ & 1.1 $\times$ 10$^{-13}$ \\ 
50\% & 7.6 $\times$ 10$^{-14}$ & 3.9 $\times$ 10$^{-14}$ & 8.6 $\times$ 10$^{-14}$ \\ 
20\% & 5.9 $\times$ 10$^{-14}$ & 2.9 $\times$ 10$^{-14}$ & 6.4 $\times$ 10$^{-14}$ \\
\hline
\hline
\end{tabular}
\label{limit_flux}
\end{table}

\subsection{Flux analysis}
To validate the aperture photometry performed and the count deblending applied, we computed the 3-24 keV (and 3-8 and 8-24 keV) band fluxes for all detected sources. We use the counts-to-flux conversion factors reported in Section~\ref{simul1}, computed using a power law model with $\Gamma$=1.8 and Galactic column density. 
We then convert the fluxes from aperture (in 20$^{\prime\prime}$) to total assuming a factor, derived from the \nustar\ point spread function, such that F$_{aperture}$/F$_{total}$=0.32. We find that this value is approximately constant across the field of view (with only a few percent variation) because the size of the PSF core is constant. Therefore, this factor can be applied to all the sources to convert the aperture counts computed from the mosaic, i.e. using the counts from different positions on the detector, to total counts. The difference of the aperture correction between FPMA and FPMB is of the order of 4\%, and would affect the flux estimates at the level of the statistical uncertainties. Moreover, this aperture correction factor is energy independent and can be applied to convert from aperture to total fluxes in the 3 bands used here.

\begin{figure}
\centering
\includegraphics[width=0.45\textwidth, angle=0]{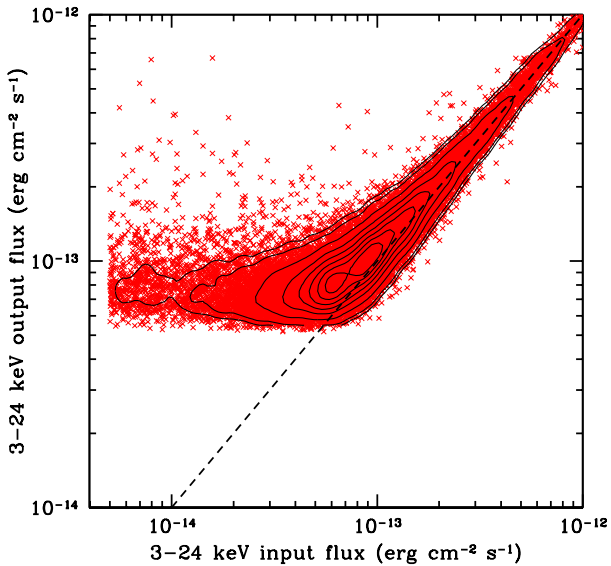}
\caption{\small Comparison between 3-24 keV input fluxes and measured total fluxes in the simulations for all the sources with DET\_ML above the 99\% reliability threshold in the 3-24 keV band. Black contours show the source density. }
\label{fxfx}
\end{figure}

In Figure~\ref{fxfx}, we present the input versus output fluxes in the 3-24 keV band for all sources above the detection threshold in all 400 simulations. The agreement at bright fluxes validates the aperture correction. The fluxes are within a factor 1.5 of the input value down to the flux corresponding to almost 90\% completeness (10$^{-13}$ \cgs). The spread of the distribution increases towards lower fluxes, becoming a factor of 2.5 wider at the flux corresponding to 50\% completeness of the survey (7$\times$10$^{-14}$ \cgs). The spread at low fluxes, in particular at the flux limit, is expected, and is due to  Eddington bias.

\section{Data Analysis and Point Source Catalog}
\label{realdata}

\subsection{Source Catalog Creation}

Once we tested and optimized the source detection procedure on simulated data and defined all the matching radii on the basis of the highest rate of detected and significant sources, we applied the same procedure to the \nustar\ COSMOS data. 
Summarizing, for each band, we first ran {\it SExtractor} using the parameters defined in Section \ref{sourcedetection} and we obtained a catalog of detected sources merging all the outputs. We then extracted 20$^{\prime\prime}$ source and background counts at the detection position and corrected these counts for contamination of neighbors. Using the deblended counts, we computed DET\_ML values for all sources. We define as detected all those sources found by {\it SExtractor} and detected above threshold those for which the DET\_ML is above the threshold.

There are 81, 61 and 32 sources above the 99\% reliability detection thresholds in the 3-24, 3-8 and 8-24 keV bands, respectively. 
These numbers agree, to within a few percent, with the numbers of detections expected from the simulations as presented in the last line of Table~\ref{detections}. We compared the number of detections to those predicted by the X-ray background synthesis models described in Ballantyne
et al. (2011), which consider several luminosity functions (Ueda et al. 2003, LaFranca et al. 2005, Aird et al. 2010) as well as the most recent one by Ueda et al. (2014). We combined the luminosity functions with the spectral model of Ballantyne (2014), and folded the results with the COSMOS sensitivity curves. The models under-predict the actual number of detected sources by 20-30\% in both the 3-8 keV and 8-24 keV bands, though it is still consistent within the uncertainties. We also compared with the number counts presented in Ajello et al. (2012) in the 15-55 keV band, properly converting the flux to the 8-24 keV band (assuming a $\Gamma$=1.8 slope). Also in this case, the predicted counts (17-34 sources) is in agreement with the observed data. A more extended analysis of the \nustar\ number counts bands is described in Harrison et al. (in prep.).

We then generated a master catalog by performing a simple positional matching (30$^{\prime\prime}$ matching radius) between the catalog of sources detected in the 3-24 keV band and in the 3-8 keV band, and then matched the resulting catalog (including matches but also unmatched sources in both bands) to the 8-24 keV band catalog. 
From the master catalog, we determine the number of significant sources (with DET\_ML above the threshold) in the total sample. Table~\ref{realdetections} lists the number of different combinations of sources above the threshold in at least one band. We include the combination of sources detected and above the threshold (labeled with capital F, S and H) and detected but below the threshold (labeled with lower case f, s and h). By summing all the possible combinations, the total number of sources above the threshold in at least one band is 91.

The counts for each detected source in a given band were computed by performing photometry in a 20$^{\prime\prime}$ radius aperture in the FPMA+B mosaic as well as in the background mosaic and then correcting the counts using the deblending code as per the simulations. If the source was either detected, but below the threshold, or undetected in a given band, we computed upper limits by extracting the counts (in the same 20$^{\prime\prime}$ radius aperture) at the position of the detected (but below threshold) source, or at the position of the significant detection in another band if the source was undetected. The method adopted for error determination is Gehrels (1986; 1$\sigma$ equivalent errors are used). For non detections, 90\% confidence upper limits were computed by following standard approach (see Narsky 2000). 

\begin{figure}
\centering
\includegraphics[width=0.45\textwidth, angle=0]{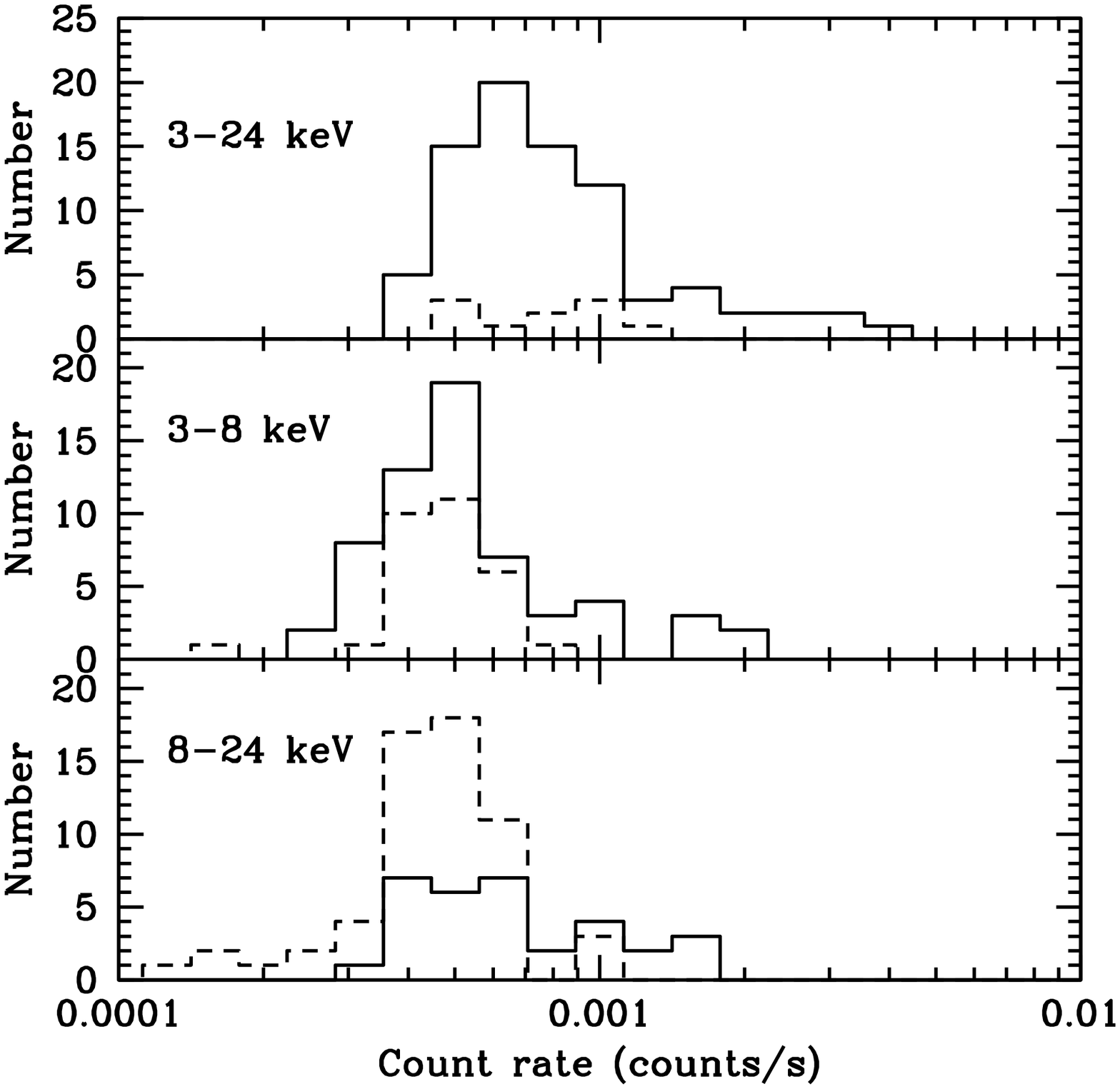}
\includegraphics[width=0.45\textwidth, angle=0]{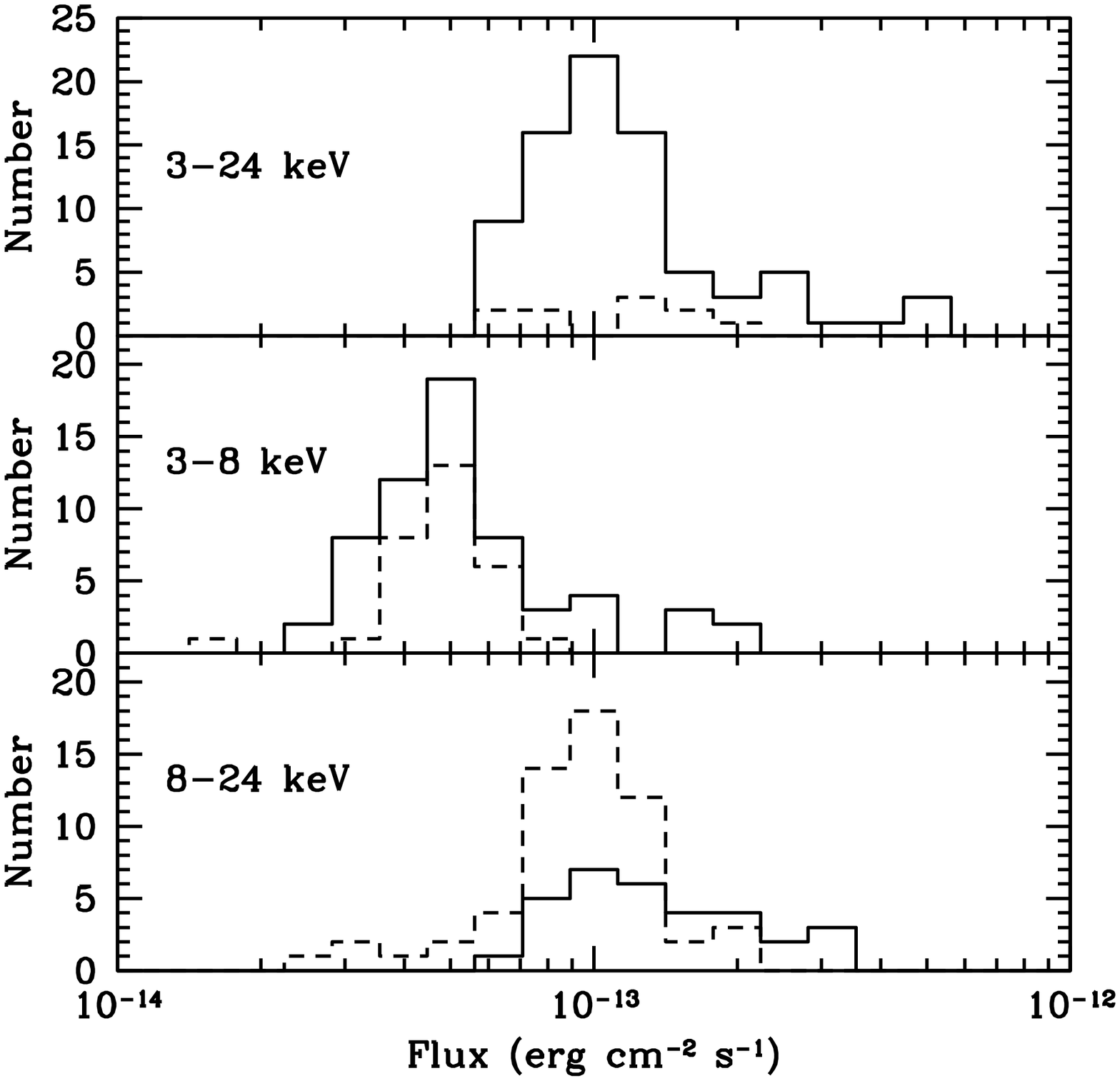}
\caption{\small Count rate (top) and flux (bottom) distributions for detected sources in the 3-24, 3-8 and 8-24 keV bands from top to bottom. Upper limits at 90\% confidence have been included for sources below threshold or undetected as dashed histogram.}
\label{flux_cr_isto}
\end{figure}

Vignetting-corrected count rates for each source are obtained by dividing the best-fit counts derived from aperture photometry for each band by the net exposure time, weighted by the vignetting at the position of each source. Total fluxes were obtained by converting the count rates, assuming a power law model with $\Gamma$=1.8 and Galactic column density, using the conversion factor in each band and applying the aperture correction factor. The energy conversion factors, and therefore the fluxes, are sensitive to the spectral shape: a flatter spectral slope ($\Gamma$=1) would produce a difference of $\sim$20\%, $<$5\% and $<$15\% in the 3-24 keV, 3-8 keV and 8-24 keV fluxes, respectively. The energy conversion factor for the 3-24 keV band depends most strongly on the spectral shape because of the wider band considered. In Figure~\ref{flux_cr_isto}, the histograms of count rates and total fluxes in each band are presented (90\% confidence upper limits have been included).

\begin{table}[]
\small
\centering
\caption{Number of Sources with DET\_ML above the defined threshold in at least one band.}
\begin{tabular}{l c }
\hline
\hline
Band &Number of sources\\
\hline

F+S+H & 23 \\
F+S+ h & 14 \\
F+S & 15\\
F+s+h &  7\\
F+s &  8\\
F+h & 4\\
F   & 2\\
F+s+H &  6\\
F+H &   2\\
f+S &  9\\
H & 1\\
\hline
\hline

\end{tabular}
\label{realdetections}

{\footnotesize F, f: 3-24 keV, S, s: 3-8 keV, H, h: 8-24 keV. Capital F, S and H refer to sources detected and above the threshold in that band, while lower case f, s, h refer to sources detected in a given band but below the detection threshold.}
\end{table}

\subsection{Match to \chandra\ and \xmm\ Catalogs}

The 91 \nustar\ detected sources were matched to both the \chandra\ (Civano et al. 2012) and XMM-COSMOS point-source catalogs (Brusa et al. 2010) to obtain lower energy counterparts and multiwavelength information. We use the nearest neighbor matching approach, as done in the simulations (Section~\ref{sourcedetection}), with a 30$^{\prime\prime}$ matching radius. Given that the \nustar\ survey overlaps also with the new \chandra\ COSMOS Legacy survey (Civano et al. in prep.), we also used the \chandra\ catalog for this project (Marchesi et al. in prep.). We applied a flux cut to both catalogs at 5$\times$10$^{-15}$ \cgs\ (2-10 keV), consistent with the limit in flux used in the simulations. At this flux, the number density of sources in the 2-10 keV band is 600 deg$^{-2}$, therefore the number of \chandra\ or \xmm\ sources found by chance in the searching area around each \nustar\ source is $<$ 0.13. All the matches are therefore very likely to be real associations.

\begin{figure}
\centering
\includegraphics[width=0.45\textwidth, angle=0]{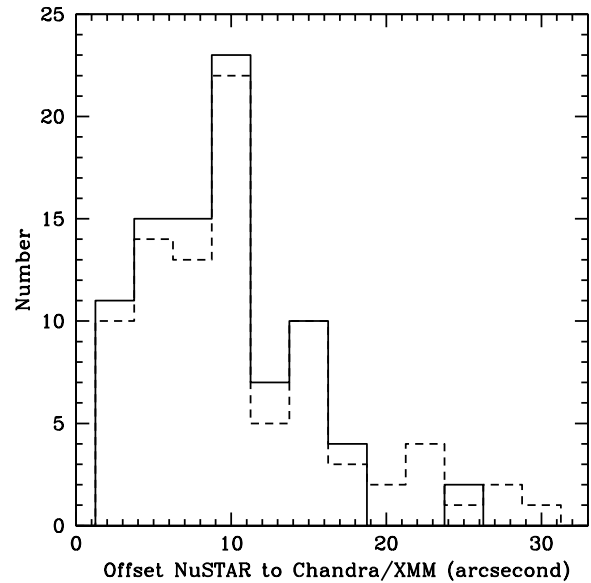}
\caption{\small Histogram of separations between the \nustar\ and  \chandra\ or \xmm\ positions of the primary (solid) and secondary (dashed) counterparts.}
\label{nu_chaxmm}
\end{figure}

The cross-correlation returned 87 matches within 30$^{\prime\prime}$. The distribution of separations between the \chandra\ or \xmm\ and the  \nustar\ positions is presented in Figure~\ref{nu_chaxmm}. The fraction of matches within 10$^{\prime\prime}$ and 20$^{\prime\prime}$ are 56\% and 97\%, respectively, both consistent with the simulations (see Figure~\ref{distances}, dashed lines). 
Among the 87 sources with a \chandra\ and/or \xmm\ counterpart, 14 have multiple matches within 30$^{\prime\prime}$. The distribution of separations considering the secondary counterpart (defining as secondary the source at larger separation) is shown in Figure~\ref{nu_chaxmm} as a dashed line. In this case, the fraction of matches within 10$^{\prime\prime}$ drops to 52\%, and to 90\% for matches within 20$^{\prime\prime}$. Two of the \nustar\ sources with a multiple match show a significant iron K$\alpha$ line in the \nustar\ spectrum at the energy as expected from the redshift of the primary \chandra\ and/or \xmm\ counterpart, therefore these sources (\nustar\ J100142+0203.8 and J100259+0220.6, source ID~181 and ID~330 in the catalog) can be securely associated with their lower energy counterpart. Of the 12 remaining \nustar\ sources with multiple lower X-ray energy counterparts, only two (\nustar\ J095845+0149.0 and J095935+0241.3, source ID~134 and ID~257) have a primary and secondary counterpart with separations and fluxes in the 2-10 keV \chandra/\xmm\ band within 3\% one of the other; therefore both primary and secondary could be considered a possible association. For ten of the 14 sources with two possible counterparts, the separation between the \nustar\ position and the secondary candidate is 30\% larger than the separation between the \nustar\ position and the primary. The flux of the secondary is also significantly fainter (50\% or more) than the primary, making the primary association stronger. Hereafter, we consider the primary match to be the \chandra\ and/or \xmm\ counterpart. We flag those \nustar\ sources with secondary counterparts, providing a supplementary catalog of matches.

\begin{figure}
\centering
\includegraphics[width=0.45\textwidth, angle=0]{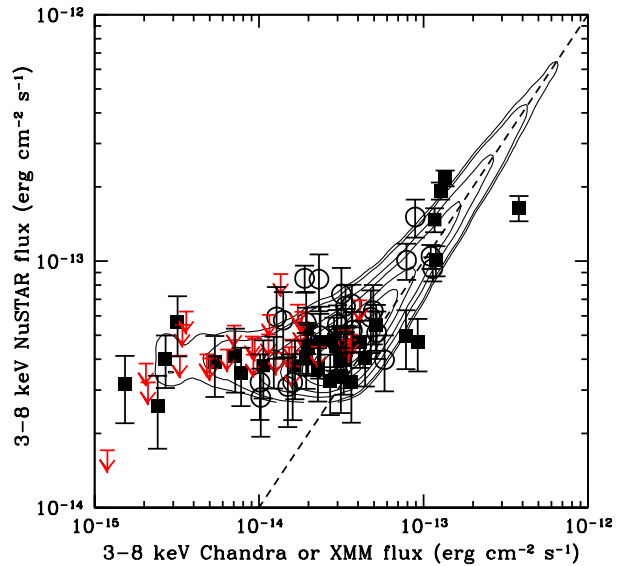}
\caption{\small Comparison between the \chandra\ (full square symbols) or \xmm\  (empty circle symbols) and the \nustar\  3-8 keV fluxes. \nustar\ 90\% upper limits are plotted in red as downward arrows. Black contours show the density of sources in the simulated sample when comparing 3-8 keV input fluxes and measured/deblended fluxes as in Figure \ref{fxfx}. }
\label{fx_nu_chaxmm}
\end{figure}

Of the 87 matched sources, 79 are associated with a \chandra\ and {\it XMM}-COSMOS source (41 to a C-COSMOS source and 38 to a \chandra\ COSMOS Legacy source), seven with a \chandra\ only source (of which four are from the new \chandra\ COSMOS Legacy survey), and only one is matched with an {\it XMM}-COSMOS source outside the \chandra\ COSMOS Legacy survey area (see also Figure \ref{mosaic}). Although the \chandra\ COSMOS Legacy survey overlaps with the {\it XMM}-COSMOS area, we consider here fluxes from the already published {\it XMM}-COSMOS catalog for the 38 sources detected in both.
All 87 sources are detected in the 2-10 keV band in at least one of the three low energy surveys. 
When considering the \chandra\ and \xmm\ sources in the central area of the \nustar\ survey that has uniform and deepest exposure, all those with 2-10 keV (\chandra\ or \xmm) fluxes larger than 10$^{-13}$ \cgs\ are also detected by \nustar. At fluxes in the range 5$\times$10$^{-14}$ -10$^{-13}$ \cgs, the fraction of  \chandra\ and \xmm\ sources detected by \nustar\ drops to 60\% and becomes lower than 10\% at fluxes below 2$\times$10$^{-14}$ \cgs. These fractions are in full agreement with the \nustar\ survey completeness presented in Table \ref{limit_flux} and Figure \ref{completeness}.
In Figure~\ref{fx_nu_chaxmm}, we compare the 3-8 keV \nustar\ fluxes with the \chandra\ or \xmm\ fluxes in the same band. We converted count rates from both the C-COSMOS and {\it XMM}-COSMOS catalogs in the 2-7 keV and 2-10 keV bands respectively into 3-8 keV fluxes, accounting for the slightly different energy range covered and the spectral model assumed here. In Figure ~\ref{fx_nu_chaxmm}, downward arrows are 90\% confidence upper limits on the \nustar\ flux . Black contours represent the locus occupied by the simulated sample when comparing input and measured fluxes in the simulations in the 3-8 keV band (as shown in Figure~\ref{fxfx} for the 3-24 keV band). The scatter around the 1:1 line is similar to that observed in the simulations, keeping in mind that flux variability could affect the real source distribution. The break observed at $\sim$10$^{-13}$ \cgs\ and the flattening of the distribution are both consistent with the simulation results and with Eddington bias affecting the sources at the flux limit. Similar behavior is found by Mullaney et al. (2015) in the ECDFS \nustar\ survey and also in the serendipitous source sample presented by Alexander et al. (2013).

\begin{figure}
\centering
\includegraphics[width=0.45\textwidth, angle=0]{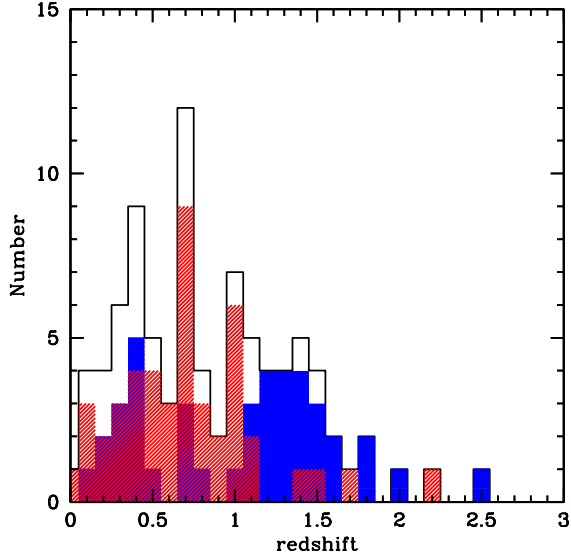}
\caption{\small Redshift distribution: black solid line= total sample; blue solid histogram = broad line and unobscured AGN; red dashed histogram= narrow line and obscured AGN.}
\label{zetadist}
\end{figure}

The redshift distribution of the 87 \nustar\ sources associated with a lower energy counterpart is presented in Figure~\ref{zetadist}.
Spectroscopic redshifts and optical spectral classifications are available for 80 matched sources and photometric redshifts (and spectral energy distribution classifications) for 87 sources from either the XMM or C-COSMOS catalogs. According to the spectroscopic classifications, the sample is equally divided between broad line AGN and narrow line AGN. For the remaining seven sources without optical spectra, the spectral energy distribution fitting suggests six are best fitted by a narrow line AGN-like template and one is best fit by a broad line AGN-like template.  Although we detect both broad line and narrow line AGN up to $z\sim$2, the mean redshift of the broad line AGN is $z\sim$1, while the narrow line AGN and narrow line AGN-like sources peak at lower redshift ($z\sim$0.6). Narrow line AGN and/or obscured sources are on average fainter (and less luminous) than broad line AGN and/or unobscured ones, and therefore are generally detected at lower redshifts, explaining the different redshift distributions and implying that the volume sampled when surveying obscured sources is smaller than that sampled by unobscured sources.

\begin{figure}
\centering
\includegraphics[width=0.45\textwidth, angle=0]{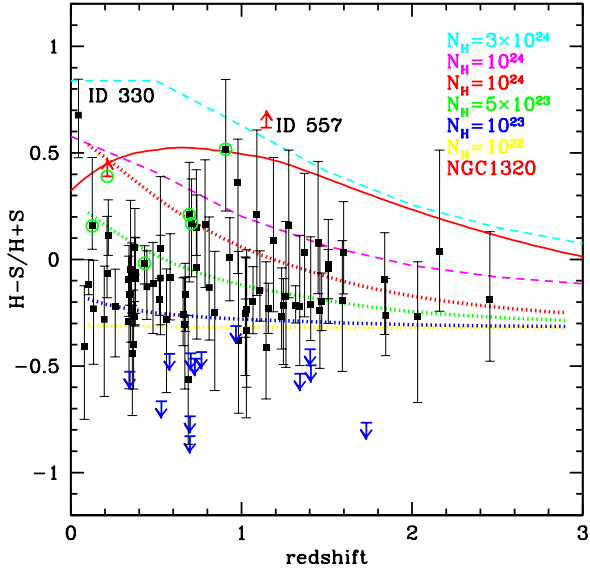}
\caption{\small Hardness ratio (with 1 $\sigma$ errors) distribution versus redshift for all the sources matched to a \chandra\ or \xmm\ counterpart. Blue downward arrows represent 1 $\sigma$ upper limits on the HR value as computed by BEHR. Red upward arrows represent lower limits. Dotted lines show power law models with $\Gamma$=1.8, Galactic column density and intrinsic column densities of 10$^{22}$ cm$^{-2}$, 10$^{23}$ cm$^{-2}$, 5$\times$10$^{23}$ cm$^{-2}$ and 10$^{24}$ cm$^{-2}$  (from bottom to top) as function of redshift. Dashed lines show the HR predicted using the more complex MYTorus model with column densities of 10$^{24}$ cm$^{-2}$  and 3$\times$10$^{24}$ cm$^{-2}$ (from bottom to top). The red solid line represents the HR evolution of NGC 1320 best spectral model from Balokovi{\'c}  et al. 2014. The green circles label the candidate obscured sources by Lanzuisi et al. (2015).}
\label{hr}
\end{figure}

We refer to Zappacosta et al. (in prep.) and Del Moro et al. (in prep.) for a detailed X-ray spectral analysis of both bright and faint sources and for the absorption distribution in the sample.
Here, to obtain a rough estimate of the obscuration level characterizing the COSMOS \nustar\ sources, we computed the hardness ratio, defined as HR=$\frac{H-S}{H+S}$, where H and S are the number of net counts in the 8-24 keV and 3-8 keV bands, respectively. We used the Bayesian Estimation of Hardness Ratios method (BEHR, Park et al. 2006) which is the most suitable tool to compute hardness ratios and uncertainties in the Poisson regime of low counts, whether the source is detected in both bands or not. The HR reported in the catalog is the mode value computed by BEHR. 
To compare the HR (in the 3-8 and 8-24 keV bands) computed above with X-ray spectral models, and to characterize the level of intrinsic obscuration, the redshift of each source needs to be taken into account. In Figure~\ref{hr}, the HR values are plotted for each source versus redshift. If the upper or lower value of the HR is at its maximum value (1 or $-$1, respectively), the HR values are considered to be lower or upper limits. The error computed with BEHR were estimated using the Gibbs sampler (a special case of the MCMC) to obtain information on the posterior distribution of the 3-8 and 8-24 keV counts and therefore on the HR (see Park et al. 2006 for more details).  The errors and the upper and lower limits on the HR are derived from the MCMC draws. The limits do not necessarily correspond to a non detection in a given band, because BEHR computes HR directly using total counts and background counts, and relies on the combined statistics of both sub-bands. 

Even though spectral complexity is likely present in these sources (see Section \ref{spectral_analysis} as an example), we compared the HRs with two sets of spectral models: a power law model with slope of $\Gamma$=1.8 and column densities 10$^{22}$ cm$^{-2}$, 10$^{23}$ cm$^{-2}$, 5$\times$10$^{23}$ cm$^{-2}$ and 10$^{24}$ cm$^{-2}$  (dotted lines, from bottom to top) and the more complex MYTorus model (Murphy \&\ Yaqoob 2009). For the latter, we assumed a uniform torus with opening angle with respect to the axis of the system fixed to 60~deg, corresponding to a covering factor of 0.5, and column densities of 10$^{24}$ cm$^{-2}$ and 3$\times$10$^{24}$ cm$^{-2}$ (dashed lines, from bottom to top). Balokovi{\'c} et al. (2014) used the latter model in spectral analysis of particular heavily obscured AGN observed with \nustar\ (NGC 1320), and found hardness ratios calculated from the best-fit models which are consistent with those considered here. As for comparison we also report in Fig. \ref{hr} the hardness ratio evolution computed for the best fit spectra model of NGC 1320 (with 4$\times$10$^{24}$ cm$^{-2}$, red solid line) from Balokovi{\'c} et al. (2014).

As shown in Figure~\ref{hr}, if we choose HR=-0.2, corresponding to the commonly adopted definition of obscured X-ray AGN (10$^{22}$ cm$^{-2}$), to divide the sample in obscured and unobscured, we find that ~50\% of the \nustar\ sources are obscured. We caution that in the 3-24 keV \nustar\ passband, the HR for a modestly obscured AGN is very close to that expected for completely unobscured AGN (HR=-0.3) and so it is difficult to estimate the reliability and statistical uncertainty in this measure of the obscured fraction from \nustar\ HR alone. A more robust and accurate measure of the fraction including modestly obscured AGN (down to 10$^{22}$ cm$^{-2}$) will require X-ray spectral analysis to measure $N_H$ directly and comparison with lower-energy \chandra\ and \xmm\ data, which will be addressed in future work (Zappacosta et al. in prep.). 
\nustar\ HR is more sensitive to higher obscuration  ($>$10$^{24}$ cm$^{-2}$), and can therefore be used to identify candidate CT sources. We refer to these sources as candidates as confirming their CT nature requires detailed analysis of their X-ray spectra to measure $N_H$ directly or other independent means, which is again beyond the scope of this paper. We compute the fraction of candidate CT sources, defined as the number of sources with HR and redshift combination above the $>10^{24}$ cm$^{-2}$ line obtained using the MYTorus model (the magenta line in Figure \ref{hr}), using the results from the BEHR MCMC analysis\footnote{We used each of the 5000 MCMC draws for each of the 87 detected source as a independent HR value for each source and we combined the draws for each source and treat them as ``new'' sample of sources to estimate the fraction of candidate CT sources.} and thus allowing for the uncertainties in the individual HR estimates. We also compute the fraction using the $10^{24}$ cm$^{-2}$ line obtained with an obscured power law model (the red dashed line in Figure \ref{hr}). The fraction of candidate CT AGN obtained is 13\%$\pm$ 3\% with the MYTorus model and 20\%$\pm$3\% with the obscured power-law model. These estimates are consistent with the fractions based simply on the best estimates (posterior mode) for HR (9\% and 19\% respectively). We note that these values correspond to the \emph{observed} fraction of CT candidates, combined over the entire luminosity and redshift range of our sample.

Two sources have lower limits on their HR: \nustar\ J100204+0238.5 (ID~557) is only detected above the threshold in the 8-24 keV band (and it is the solo source with only an 8-24 keV band detection in the whole sample) while source \nustar\ J100229+0249.0 (ID~249) is also detected above the threshold in the 3-24 keV band. Their HRs suggest high levels of obscuration, above 5$\times$10$^{23}$ cm$^{-2}$. Source ID~249 is also part of a sample of candidate highly obscured AGN from the XMM and C-COSMOS spectral analysis (Lanzuisi et al. 2015). 
A total of six sources detected by \nustar, highlighted in green in Figure~\ref{hr} (\nustar\ ID 107, 249, 299, 129, 181, 216), are identified as candidate obscured AGN by the same Lanzuisi et al. (2015) spectral analysis. For all of them, the \nustar\ HR suggests columns densities exceeding 5$\times$10$^{23}$ cm$^{-2}$, considering also the 1$\sigma$ error bars. 


Two sources (ID~330 and ID~557, labelled in the figure) have HR$>$0.5 strongly suggesting obscuration exceeding 10$^{24}$ cm$^{-2}$ with both an obscured power law model and the MYTorus model. Both of these are candidate CT AGN. A more detailed analysis of ID~330 is presented in Section~\ref{spectral_analysis}.

\begin{figure}
\centering
\includegraphics[width=0.45\textwidth, angle=0]{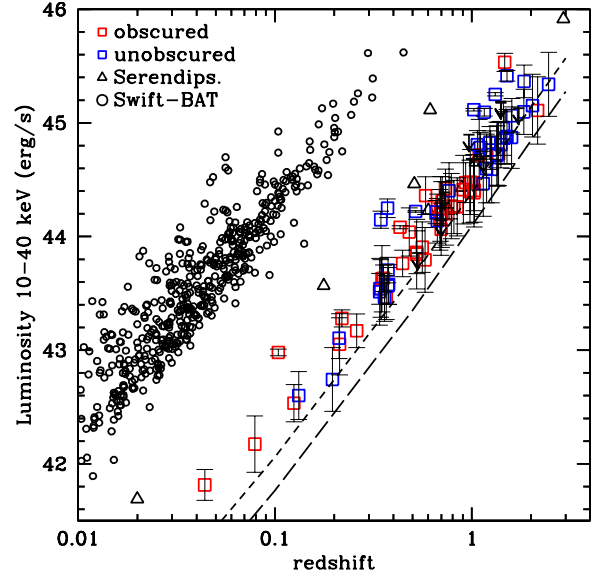}
\caption{\small 10-40 keV rest frame luminosity versus redshift for COSMOS \nustar\ sources (square symbols): in blue the sources classified as broad line or unobscured AGN and in red the sources classified as narrow line or obscured AGN. COSMOS upper limit are presented as downward black arrows. Short and long dashed lines represent the flux limit of the COSMOS survey (at 20\% completeness) and of the ECDFS survey. The serendipitous \nustar\ sources presented in Alexander et al. (2013) and {\it Swift}-BAT 70 months sample (Baumgartner et al. 2013) are shown as triangles and circles, respectively.}
\label{lxz}
\end{figure}

\nustar\ fluxes in the 3-24 keV band, where we have the highest number of detections, have been converted into 10-40 keV rest frame luminosities, assuming a power law model with $\Gamma$=1.8 and a standard k-correction of $(1+z)^{(\Gamma -2)}$ to take into account the different bandpasses. The luminosities here are not corrected for absorption, although, the 10-40 keV band is not sensitive to obscuration up to columns of $>$ few$\times$ 10$^{24}$ cm$^{-2}$. In Figure \ref{lxz}, the X-ray luminosity for the \nustar\ COSMOS sources is plotted versus redshift. Upper limits for eight sources not detected in the 3-24 keV band have been included as downward arrows. The COSMOS sample is compared here with the {\it Swift}-BAT 70 month all-sky survey sample (Baumgartner et al. 2013). The flux limit of the ECDFS survey (Mullaney et al. 2015) is also presented as a long-dashed line. The COSMOS \nustar\ survey reaches luminosities two orders of magnitude fainter than the {\it Swift}-BAT sample and extends to significantly higher redshift. Broad line and unobscured sources (blue squares) have on average higher luminosities, while the faint end of the luminosity distribution is dominated by narrow line or obscured AGN. Given the large area covered by the COSMOS survey, we are able to sample also rare sources at very low redshift and faint luminosity, such as source ID 330, a spiral galaxy at $z=$~0.044 (see Section~\ref{spectral_analysis}).

\subsection{X-ray to optical properties}
\label{xo}

To further study the nature of the \nustar\ sources, we compare the X-ray fluxes to the optical magnitudes of their counterparts.
The $X/O$ ratio (Maccacaro et al. 1988) is defined as \hbox{$X/O = \log(f_X/f_{opt}) = \log(f_X) + C + m_{opt}/2.5$}, where $f_X$ is the X-ray flux in a given energy range, $m_{opt}$ is the magnitude at the chosen optical band, and $C$ is a constant which depends on the specific filter used in the optical observations. Usually, the $r$- or $i$-band flux is used (e.g., Brandt \& Hasinger 2005). Originally, a soft X-ray flux was used for this relation, and the majority of luminous spectroscopically identified AGNs in the {\it Einstein} and {\it ASCA} X-ray satellite surveys were characterized by  $X/O$ = 0 $\pm$ 1 (Schmidt et al. 1998, Stocke et al. 1991, Akiyama et al. 2000, Lehmann et al. 2001). Harder X-ray surveys, performed with \chandra\ and \xmm, found that a large number of X-ray detected sources have high ($>$ 10) $X/O$ values, and later studies showed that high $X/O$ is associated with large obscuration (Hornschemeier et al. 2001, Alexander et al. 2001, Fiore et al. 2003, Brusa et al. 2007, Perola et al. 2004, Civano et al. 2005, Eckart et al. 2006, Cocchia et al. 2007, Laird et al. 2009, Xue et al. 2011). High $X/O$ sources are extreme in that their optical magnitude is faint due to a combination of high redshift and/or obscuration. At low $X/O$, the optical emission is dominated by the host galaxy. Given the correlation of $X/O$ with redshift, sources with low $X/O$ are also typically at low redshift. In \chandra\ and \xmm\ surveys, low $X/O$ sources have been dubbed optically dull or X-ray Bright Optically Normal Galaxies (XBONGs). Several studies have shown that these XBONGs could harbor highly obscured AGN (Comastri et al. 2002, Civano et al. 2007), but so far no clear case has been reported.

Figure~\ref{xo} presents the $i$-band magnitude plotted against the 3-8 keV (left) and 8-24 keV (right) fluxes for all \nustar\ detected sources. For both bands, the $X/O$ = $\pm$1 locus (yellow area) has been defined using $C$ = 5.91, computed taking into account the width of the $i$-band filters in COSMOS (Subaru, CFHT, or for bright sources SDSS). The locus takes into account the spectral slope used to compute the X-ray fluxes ($\Gamma$=1.8). The long dashed lines represent the region including 90\% of the AGN population as derived in the 2-10 keV band in the C-COSMOS survey (Civano et al. 2012). The four sources not matched to a \chandra\ or \xmm\ counterpart are plotted as upper limits with $i <$27 (see Section~\ref{unmatched}).
Even though this is the first time such a locus is presented above 10 keV, the agreement between detections and the AGN locus is remarkable. About 10\% of the sources in both the 3-8 and 8-24 keV bands lie outside the locus, consistent with what was found in the \chandra\ 2-10 keV band (see e.g., Civano et al. 2012). Flux upper limits are consistent with the locus moving to the left of the plot, and could increase the number of sources at high $X/O$. 
It is interesting that the two sources with HR$>$0.5 (starred symbols in Figure~\ref{xo}) are located at extremes of the diagram, one with high $X/O$ (ID~557) and $z>$1, the other at very low redshift and very low $X/O$. Both are candidate highly obscured and perhaps CT AGN (see Section~\ref{spectral_analysis}).



\begin{figure}
\centering
\includegraphics[width=0.45\textwidth, angle=0]{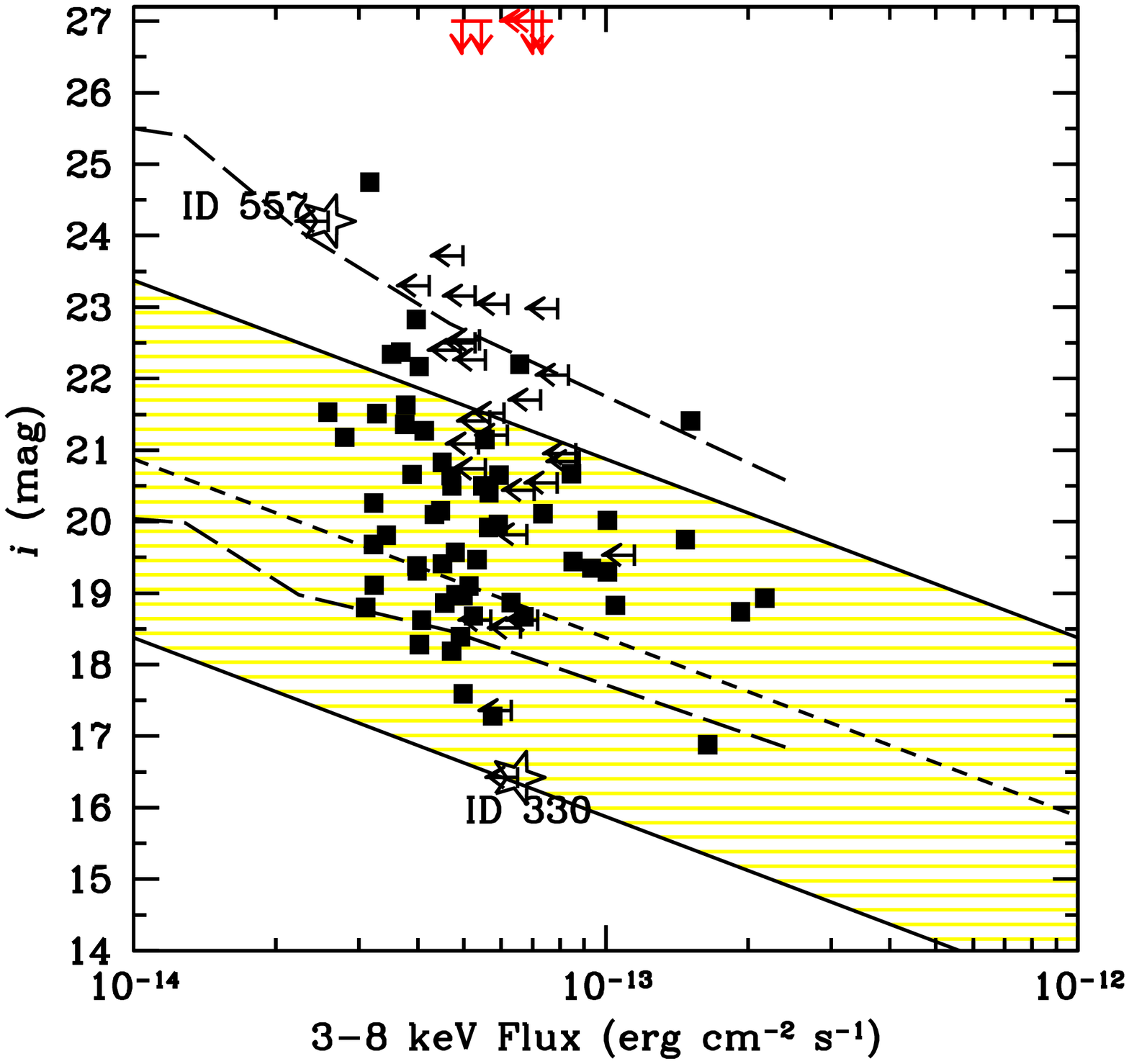}
\includegraphics[width=0.45\textwidth, angle=0]{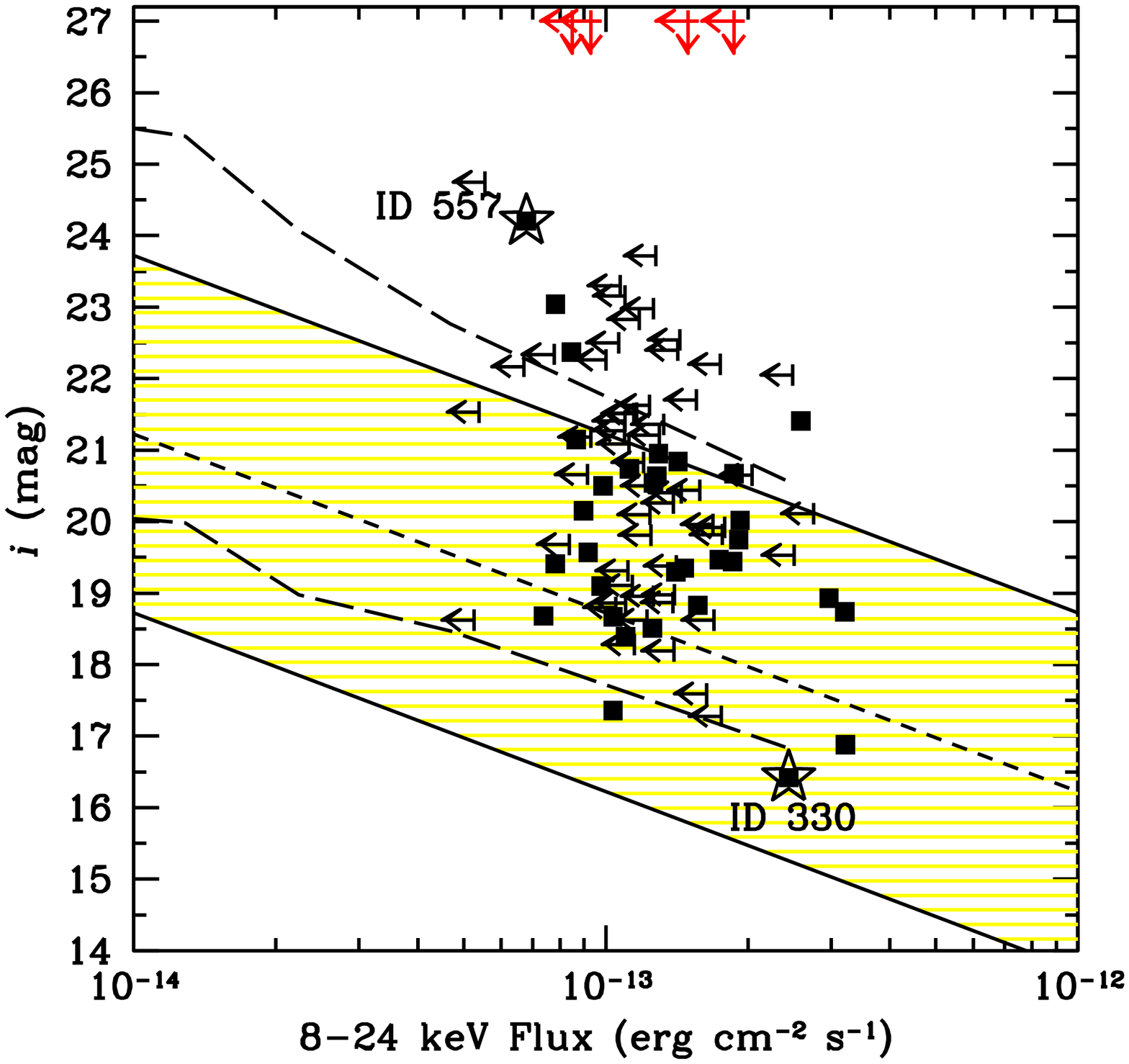}
\caption{\small X-ray flux (soft-top, hard-bottom) versus the $i$-band magnitude. The yellow shaded region represents the classic locus of AGNs along the correlation $X/O$ = 0$\pm$ 1.  The long-dashed curves represent the locus occupied in the 2-10 keV band by the C-COSMOS sources. The star symbols represent source ID~330 and ID~557.}
\label{xo}
\end{figure}

\subsection{\nustar\ sources without a \chandra\ or \xmm\ match}
\label{unmatched}

As mentioned in the Section 1, lower energy X-ray missions like \chandra\ and \xmm\ are not sensitive to very obscured AGN at redshift $z<$1.5. Therefore, any 8-24 keV detected \nustar\ source not associated with a low energy counterpart could harbor a CT AGN. Four sources (\nustar\  J100047+0139.2, J095820+0149.3, J095831+0150.1 and J095930+0250.1; ID 111, 135, 141, 245 in our catalog) are not matched with a \chandra\ or \xmm\ point source within a 30$^{\prime\prime}$ radius. 
Given the 99\% reliability cut applied to the sample, we expect $\sim$2 spurious sources in the sample out of 91 detected sources (summing the number of spurious sources expected in each band). The two spurious detections could be found among the sources with a lower energy counterpart or those without a counterpart.
The probability of spurious associations is determined by the number density of sources above the 2-10 keV flux limit used for the matching, which is 600 deg$^{-2}$ . Using this density, for each source we calculate the probability of finding a \chandra\ or \xmm\ source by chance at a distance less than the observed separations between \nustar\ sources and their primary \chandra\ or \xmm\ counterparts (see Figure \ref{nu_chaxmm}). Over the full population, we expect a total of between 1 and 2 of the associations to be due to random chance.

Source \nustar\ J100047+0139.2 (ID~111) is coincident with an extended \chandra\ and \xmm\  source centered at RA=10:00:45.55 and Dec=+01:39:26.1 with redshift $z$= 0.220 and X-ray flux F$_{0.5-2 keV}$=1.6$\times$10$^{-13}$ \cgs\ (Finoguenov et al. 2007, George et al. 2011, Kettula et al. 2013). Its strong detection in the 3-8 keV band (and no detection in the 8-24 keV band) with a flux of 4.5$\times$10$^{-14}$ \cgs\ (in agreement with the flux in the same band measured in the \xmm\ spectrum within the error bars) suggests that the \nustar\ emission also arises from a soft extended source, likely a galaxy cluster. Spectral analysis of this source, including both \chandra\ and \xmm\ data, will be presented in Wik et al. (in prep.).

\nustar\ J095820+0149.3 (ID~135) is significantly detected in the 3-8 keV band and just below the DET\_ML threshold in the 3-24 keV band.
Source \nustar\ J095831+0150.1 (ID~141) is significantly detected in the  3-24 keV band and just below the threshold in the other two bands. 
The last source, \nustar\ J095930+0250.1 (ID~245), is significantly detected in the 3-24 keV band and just below the threshold in the 8-24 keV band.

\begin{figure}
\centering
\fbox{\includegraphics[width=0.45\textwidth, angle=0]{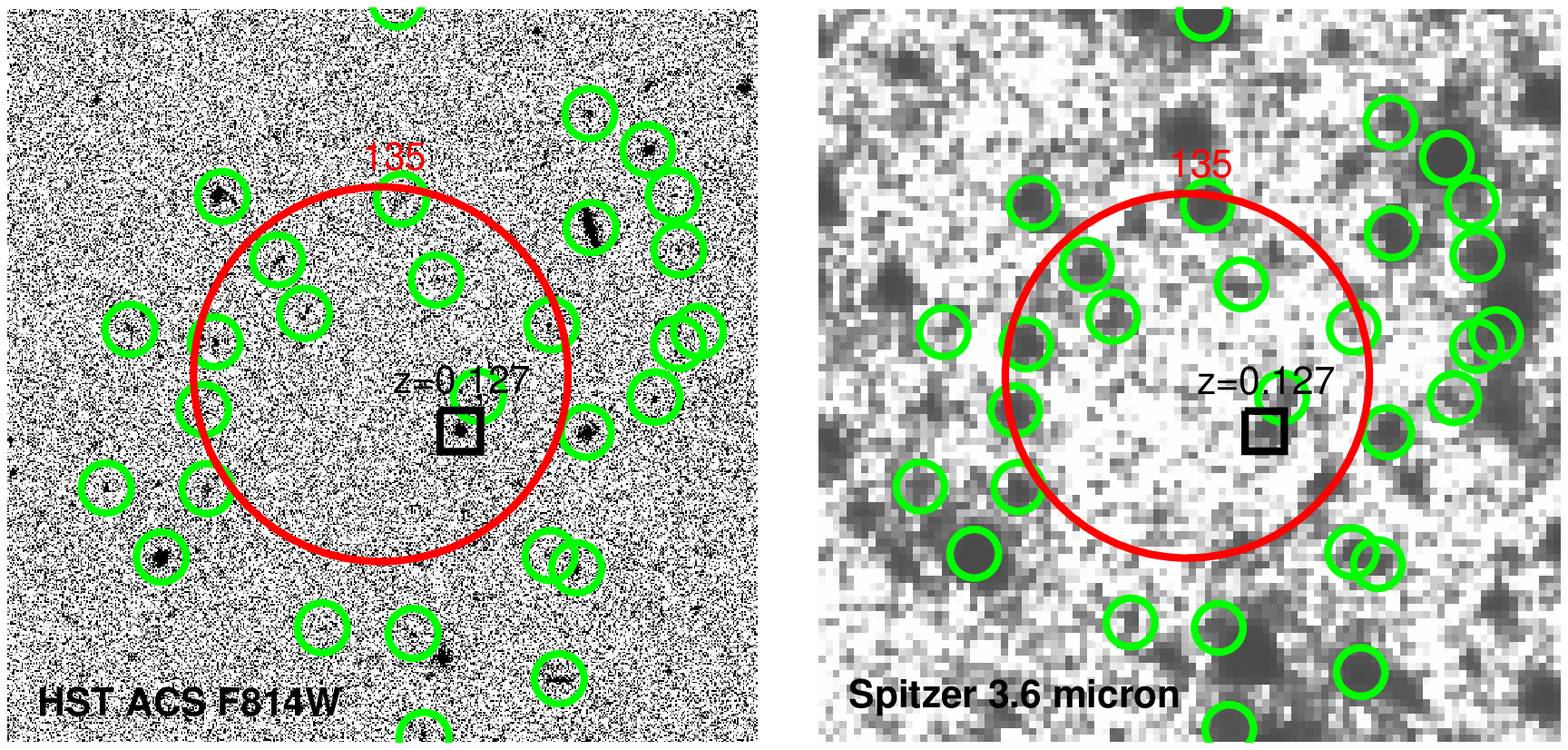}}
\fbox{\includegraphics[width=0.45\textwidth, angle=0]{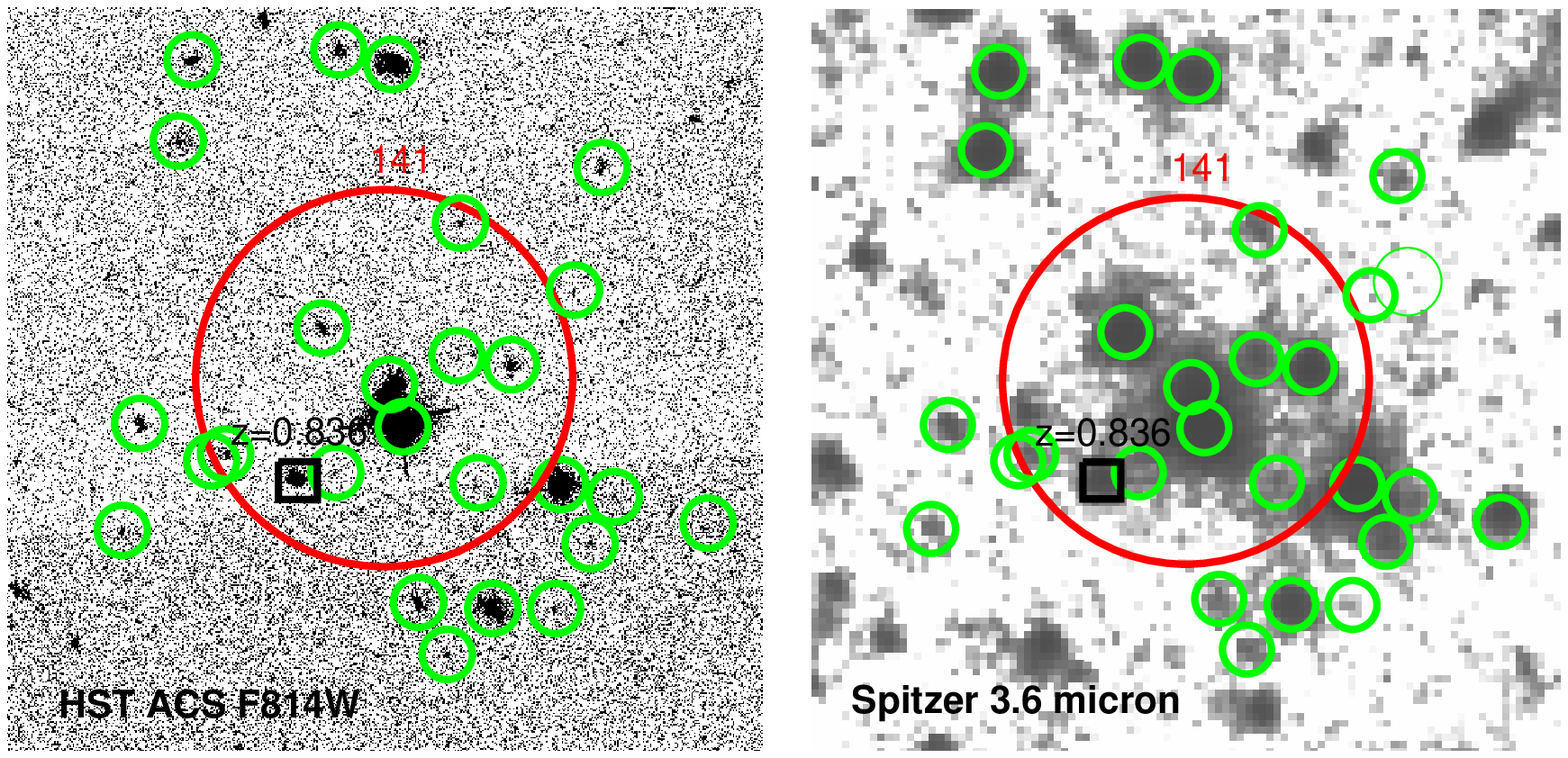}}
\fbox{\includegraphics[width=0.45\textwidth, angle=0]{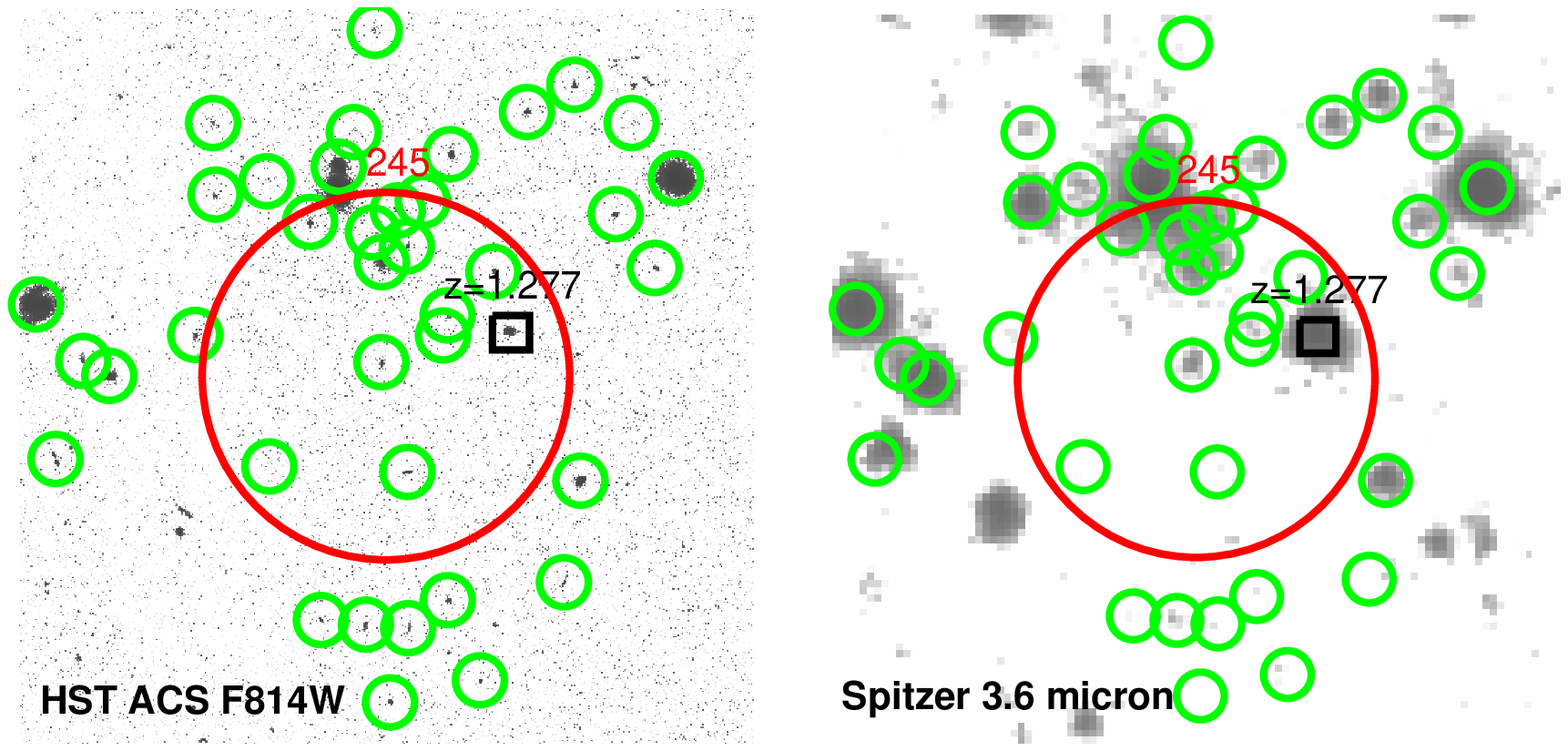}}
\caption{\small Postage-stamp images of the 30$^{\prime\prime} \times$30$^{\prime\prime}$ sky around \nustar\ sources ID 135, 141 and 245 (from top to bottom). For each source the HST ACS F814W and {\it Spitzer} IRAC (3.6 $\mu$m) images are presented. Red circles = \nustar\ position; green circles = optical catalog sources; black squares = sources with spectroscopic identification.}
\label{undetected}
\end{figure}

Figure~\ref{undetected} presents the {\it Hubble} ACS F814W and {\it Spitzer} IRAC (3.6 $\mu$m) 30$^{\prime\prime} \times$30$^{\prime\prime}$ cutouts around the \nustar\ positions for the three unmatched sources, excluding the extended one. Given their relatively bright X-ray fluxes (see red arrows in Figure~\ref{xo}), their optical infrared counterparts could be any of the objects labelled in Figure~\ref{undetected}. 
By matching the \nustar\ position of sources ID 135, 141, 245 with the COSMOS photometric catalog (Ilbert et al. 2009), we find that within 30$^{\prime\prime}$ there are about 200 detected sources, 
of which about $\sim$30 have an $i$-band magnitude that is in the same range $i$=16-25 as the 87 sources matched to a \chandra\ or \xmm\ counterpart (see Figure~\ref{xo}). 

 There are two extended optical sources, neither with a spectroscopic redshift, and a bright star, a few arcseconds away from the \nustar\ position of source ID~141. A source with a spectroscopic redshift of $z$=0.836, identified as a galaxy is detected 15$^{\prime\prime}$ from source ID~141 position.   
For source ID 245, two bright spectroscopically identified sources are detected at separations of 11$^{\prime\prime}$ and 15$^{\prime\prime}$, respectively, one at $z$=1.277 and the other at $z$=0.358. Both of these are classified as galaxies. The one at $z$=1.277 is also identified as an AGN using the infrared selection criterion of Donley et al. (2012). The catalog of IR selected AGN of Donley et al. (2012) includes $\sim$1500 sources, detected with S/N$>$3 in all four {\it Spitzer}-IRAC channels over 2 deg$^2$. \chandra\ stacking analysis 
of the individually non-detected AGN candidates leads to a hard X-ray signal indicative of heavily obscured to mildly CT obscuration. The number of expected IR selected AGN in a circle of 11$^{\prime\prime}$ radius is 0.02. Therefore the proximity of the IR AGN to source ID 245 together with its redshift make it the likely counterpart of the \nustar\ source. The measured 12\micron\ luminosity of the IR AGN ($\sim$10$^{45}$ erg/s), and the 2-10 keV luminosity as derived from the \nustar\ flux ($\sim$6$\times$10$^{44}$ erg/s) follows the relation between the observed infrared and intrinsic X-ray luminosities of AGN (e.g. Gandhi et al. 2009; Fiore et al. 2009). The \chandra\ upper limit on the 2-10 keV luminosity is instead significantly fainter than the intrinsic value estimated from \nustar\ ($<$2$\times$10$^{43}$ erg/s), implying very high obscuration. 

For source ID 135, only one spectroscopically identified galaxy at $z$=0.127 is detected $\sim$9$^{\prime\prime}$ away the \nustar\ position therefore a possible counterpart given the separation distribution as shown in Figure \ref{nu_chaxmm}, and its optical magnitude of  $i_{AB}$=22.5, which places it within the AGN locus in Figure \ref{xo}, top panel.


As a general conclusion, assuming these three unmatched sources are all narrow lines AGN with a mean redshift of $z$=0.6, their X-ray luminosities in the 10-40 keV band would be 10$^{44}$ \lum\ or higher. 
Due to their relatively bright X-ray fluxes, and the significantly deeper flux limits in the 2-10 keV band of the \chandra\ COSMOS Legacy survey ($\sim$3$\times$10$^{-15}$ \cgs), X-ray variability could explain their lack of low energy counterparts. 

X-ray flux variability on timescales from hours to years by factors 10-100 is uncommon, although such events have been detected in the past (e.g., Ulrich et al. 1997, Uttley et al. 2002, McHardy 2013, Lanzuisi et al. 2014). Source ID 141 and 245 are only detected in the full \nustar\ band, so a direct comparison with the COSMOS 2-10 keV flux limit cannot be made. Source ID 135 is only detected in the 3-8 keV band with a flux a factor 10 brighter than the \chandra\ limit. This could be explained by variability. 


\section{Revealing highly obscured sources: the case of source ID 330}
\label{spectral_analysis}

While spectral analysis is the best tool to measure intrinsic source properties (spectral index, column density, luminosity), using the hardness ratio it is possible to obtain a rough measurement of obscuration, although a simply hardness ratio selection does not return a complete sample of obscured sources. From the analysis of the hardness ratio of the 87 \nustar\ sources shown in Figure~\ref{hr}, two sources exhibit an extreme hardness ratio of HR$>$0.5, including error-bars. 
Source ID 557 is detected only in the 8-24 keV band with 50 net counts. The low number of counts and the proximity to a bright \nustar\ source 65$^{\prime\prime}$ away 
does not allow us to perform an accurate spectral extraction and reliable modeling. Therefore, here we present spectral analysis and properties of only source ID 330.

Source ID 330 (XMMID 5371 and also \chandra\ Legacy ID 1791), with HR=0.68$^{+0.17}_{-0.20}$ at $z$=0.044 (Falco et al. 1999), was observed in two different fields (MOS113 and MOS115) for a total net exposure time of 53~ks for each module. 

We extracted the counts in FPMA and FPMB from circular regions centered on the \nustar\ source position obtained from the detection in the FPMA+B mosaic. The radius of the extraction region was appropriately chosen to jointly maximize the total (i.e. from all the observations pertaining to the source) signal-to-noise ratio (SNR) and the number of net counts inside the aperture. For the SNR calculation and the net counts we used background counts from the background maps extracted inside the same source extraction region. 
The net collected counts for FPMA and FPMB are respectively 124 and 108 in a 55$^{\prime\prime}$ radius aperture in the 3-24 keV band. The spectral products were extracted using the NuSTARDAS task {\it nuproduct}, which also creates the corresponding response matrix and ancillary files. Simulated high-statistic background spectra were created using the {\it nuskybgd} routine. The spectral products (spectra, response matrices and ancillary files) were summed to obtain a single set of files for FPMA and FPMB. Each spectrum has been grouped at a minimum of 1 net count (i.e. background subtracted) per bin.

We performed joint modeling of the \nustar\ spectra with the \xmm\ - pn and MOS1-2 ($\sim$40~ks and $\sim$166~ks, for 83 and 112 net counts respectively) and \chandra-ACIS-I (148~ks, for 185 net counts; Marchesi et al. in prep.) spectra in the broad energy band 0.5-24~keV. The spectral modeling was performed in XSPEC version 12.8.2, using Cash statistics implemented with direct background subtraction, with a spectral binning of 5 net-counts (i.e. background-subtracted). The results of the spectral analysis reported in the following are listed in Table \ref{ID330} (1$\sigma$ errors on parameters are reported in the table for simplicity; upper limits are 90\% confidence level). 

First, we fit the spectrum using a simple absorbed power-law model (wabs$\times$pow). To account for possible variation of the source flux among observations taken at different times we added to this model a multiplicative constant which, at the end of the fitting procedure, we left free to vary to adjust the level of normalization of the spectra from each detector. The spectral slope is very flat with $\Gamma$= 1.1 but an upper limit on the obscuration is obtained and strong residuals are present at both low and high energy, as indicated from the relatively high Cstat/dof found (176.4/98), and at the position of the Fe-K$\alpha$ emission line.

\begin{figure}
\centering
\includegraphics[width=0.3\textwidth, angle=-90]{Fig18a_pr2.ps}
\includegraphics[width=0.35\textwidth]{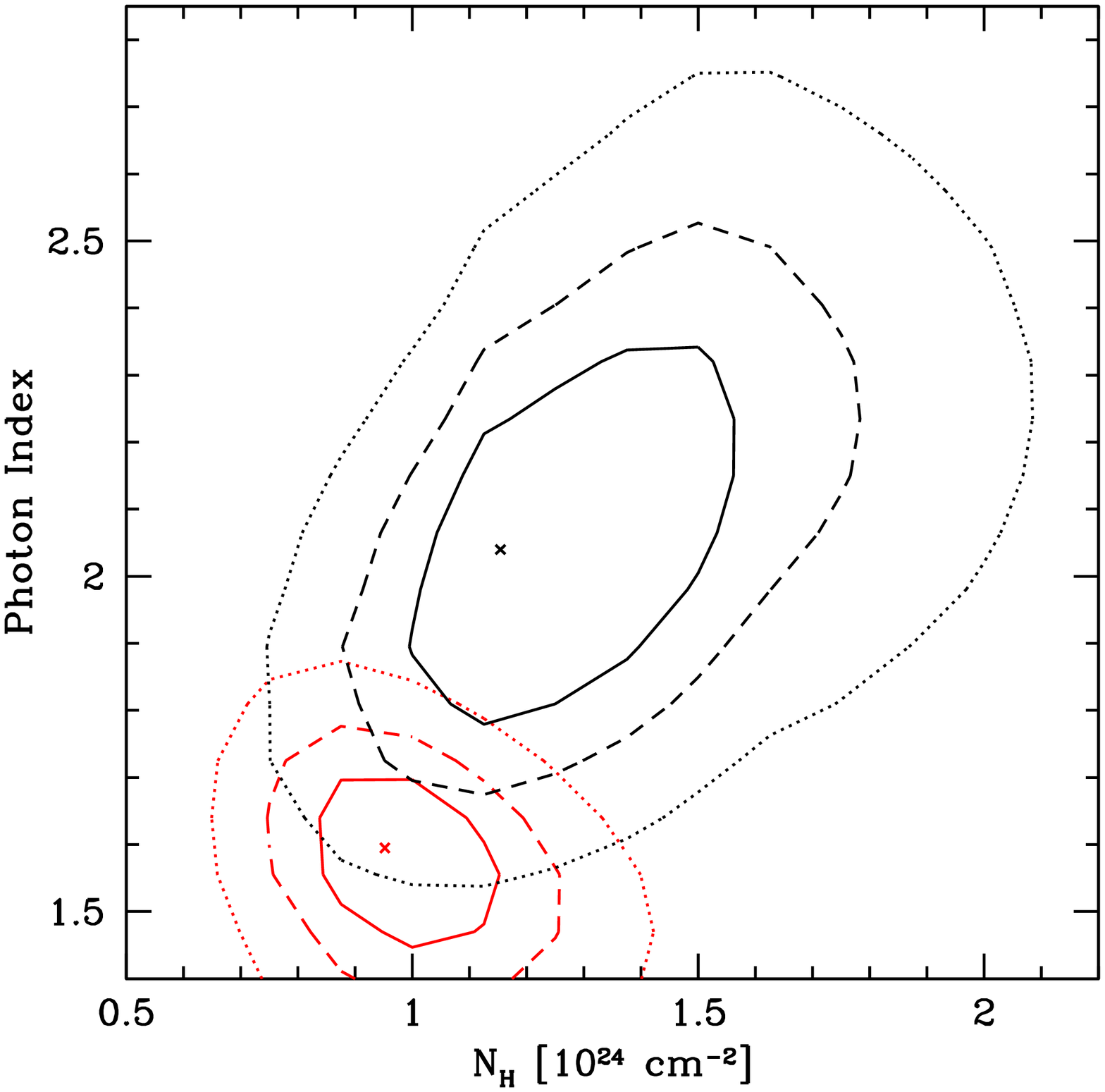}
\fbox{\includegraphics[width=0.35\textwidth]{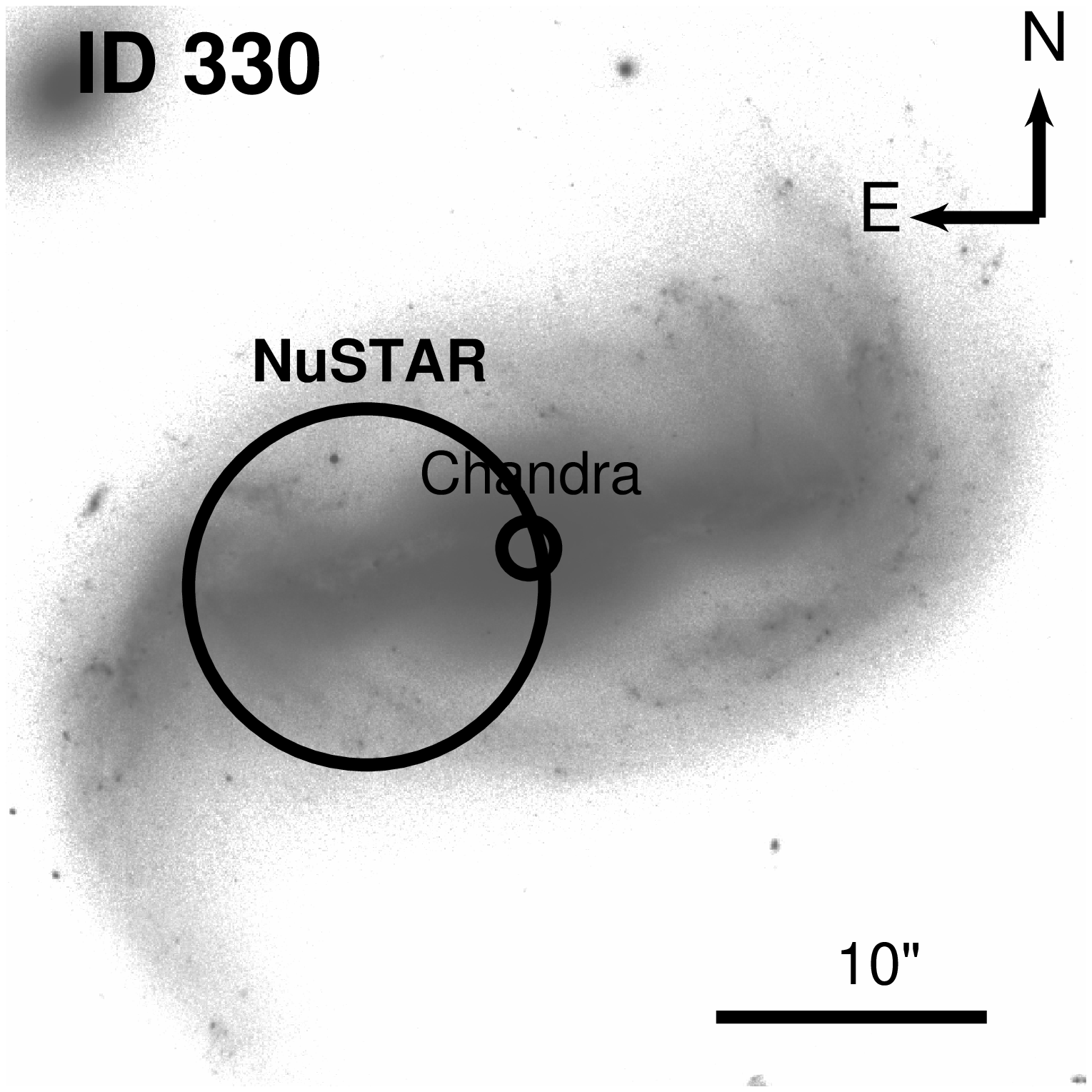}}
\caption{\small Source ID 330. {\it Top:} X-ray spectra, best-fit model and residuals for source 330. The model used is the baseline model (a simple absorbed power-law plus {\tt pexrav} component and Fe~K$\alpha$ line). Black, red, green, blue and cyan spectra refer to respectively \chandra, \xmm\ PN and MOS1-2, \nustar\ FPMA and FPMB. {\it Middle:} Confidence contours for column density and photon index for MYTorus (red) and Brigthman \& Nandra torus model (black). Solid, dashed and dotted contours refer to 68.3\%, 90\% and 99\% confidence level. {\it Bottom:} {\it HST}/ACS F814W image of the optical counterpart of source ID 330. The \nustar\ and \chandra\ positions are shown. }
\label{id330}
\end{figure}

We therefore added to the absorbed power-law model a reflection component ({\tt pexrav} in XSPEC, Magdziarz \& Zdziarski 1995) and a Gaussian line to best fit the Fe-K$\alpha$ line at the redshift of the source. To this model, which we refer to as ``baseline'', we also added a low-energy power-law component to parametrize residual flux at lower energies resulting from any scattered component. 
The spectrum of ID~330, shown in Fig.~\ref{id330} (top) along with the best-fit model and residuals, shows heavy absorption, a prominent Fe~K$\alpha$ line and a significant scattered component at energies $<$4~keV. The Cstat/dof obtained with the baseline model is significantly better than the one reported for the first model (91.3/99 versus 176.4/98), indicating a better fit.

According to the baseline model, this source would be classified as CT with $N_{\rm H}=1.6 \times 10^{24}$ cm$^{-2}$ and $\Gamma= 2.1$. The scattered power-law has a slope consistent with the primary component, and a normalization which is $\sim1\%$ of the primary emission. This suggests that the scattered component is due to a leaky absorber. 
The iron line strength of $EW_{FeK\alpha}\simeq$249~eV, even if not as extreme as expected for a leaky absorber, is consistent with the findings for CT sources in the local universe (Guainazzi, Matt \& Perola, 2005; Fukazawa et al. 2011). We therefore tied both slopes. This leaves column density and spectral slope unchanged but improved the constraints on both parameters (see Table \ref{ID330}). The \nustar\ flux in the 3-8~keV band of $2.4\times10^{-14}$ \cgs is consistent with the \chandra\ flux, and lower but within a factor 1.4 of the \xmm\ flux. 

The high value of the column density (at CT levels) estimated for this source requires a more careful analysis, and the estimation of the true column density of absorbing material by properly accounting for Compton scattering and geometry of the absorbing medium.
Therefore we tried two Monte Carlo models which self-consistently deal with absorption and scattering in a CT medium with toroidal geometry. The first model is MYTorus (Murphy \& Yaqoob 2009) and the second is the torus model from Brightman \& Nandra (2011; BNTORUS in Table \ref{ID330}), which approximates the torus as a sphere with a biconical opening. For both models, we used a torus configuration with inclination of 85~deg (i.e. almost edge-on), 60~deg opening angle and assuming a primary power-law component with an additional power-law to model the low energy excess. For both models we obtained highly uncertain values for the primary power-law photon index component which could not be constrained given the limited range of tabulated values in the models. Therefore we tied the photon indices of the scattered and primary components.
In the case of MYTorus we obtain $N_{\rm H}=10^{24}$ cm$^{-2}$ (along the line of sight) and $\Gamma$= 1.6. The equivalent width of the Fe~K$\alpha$ line is 316~eV.
For the BNTORUS model, we obtain $N_{\rm H}=1.2 \times 10^{24}$ cm$^{-2}$ and $\Gamma$= 2.0. In this model, we also estimate the torus opening angle to be 77~deg. Both models require column densities of the order of $10^{24}$ cm$^{-2}$, and give results consistent within their 1$\sigma$ errors, with MYTorus estimating slightly lower values for both parameters as shown from the confidence contours reported in Fig.~\ref{id330}, middle panel. 
The intrinsic 2-10 and 10-40 keV luminosities of ID 330 are 2.8-5.9$\times 10^{42}$ \lum\ and $\sim$5$\times 10^{42}$ \lum, consistent between both models\footnote{The 2-10 keV luminosity computed using MYTorus is lower than the one using BNTORUS because in MYTorus it is not possible to put the column density to zero, and the minimum is 10$^{22}$ cm$^{-2}$ which still affect the 2-10 keV luminosity range.}. Overall, the quality of the fit is equally good (Cstat/dof$\sim$1) between the baseline, MYTorus and BNTORUS and all of these models return a consistent value of the spectral parameters.

In the optical, this source is a barred, isolated, spiral galaxy with an inner ring (Figure~\ref{id330}, bottom panel) and a substantial bulge as classified by Hernandez-Toledo et al. (2010). 
Contrary to what was found for the obscured ECDFS source $J033202-274650$ (see also Civano et al. 2005), the $X/O$ of ID~330 is very low, as already discussed in Section~\ref{xo}, using both 3-8 keV \nustar\ and/or 2-10 keV \chandra\ fluxes (star symbol in Figure~\ref{xo}). Overall, having a low $X/O$, source ID~330 could then be classified as an XBONG at least using fluxes below 10 keV, while using higher energies fluxes could have been recognized as an AGN, being at the edge of the AGN locus in the 8-24 keV band (star symbol in Figure~\ref{xo}, bottom panel). Moreover, the \chandra\ hardness ratio (HR=0.24$\pm$0.15, Marchesi et al. in prep.) does not suggest strong obscuration. 

The X-ray spectral analysis performed with \chandra\ and/or \xmm\ data fitted separately (see Mainieri et al. 2007; Lanuzuisi et al. 2013) or together with an absorbed power law model shows a flat spectral slope with $\Gamma$=1.54$\pm^{+0.14}_{-0.12}$ and relatively low column density of $N_{\rm H}<5.4 \times 10^{20}$ cm$^{-2}$, in agreement with the hardness ratio. The intrinsic 2-10 keV luminosity derived from the \chandra\ and \xmm\ data fit alone is $\sim$1.4$\times 10^{41}$ \lum, 30 times lower than the intrinsic one measured with joint fitting performed using \nustar\ as well.

As an independent check of the NuSTAR-derived X-ray power, we can predict the intrinsic (absorption-corrected) luminosity of the source based upon known correlations between the observed mid-infrared and intrinsic X-ray luminosities of AGN (e.g. Gandhi et al. 2009; Fiore et al. 2009). For this comparison, the observed mid-infrared power is derived from results of the {\it WISE} mission (Wright et al. 2010). We use the publicly-available profile-fitting magnitudes in the AllWISE database to obtain $L_{\rm 12 \mu m}$ = 2.3 $\pm$0.2 $\times$ 10$^{43}$ \lum\ (assuming standard WISE zeropoints; Jarrett et al. 2011). Converting $L_{\rm 12 \mu m}$ to an intrinsic X-ray power ($L_{2-10 \rm \, keV}$) using the results by Gandhi et al. yields $L_{2-10 \rm \, keV}$ = 9.0$\pm$1.4 $\times$ 10$^{42}$ erg/s (the quoted 68\% confidence includes observed and systematic errors on the WISE flux, as well as the correlation scatter), which is nearly two orders of magnitude higher than the 2-10 keV luminosity measured by \chandra\ and \xmm\ only, but in agreement with the one derived performing the joint fitting. 
Given that for a typical AGN power-law photon index of $\Gamma$=1.9, L$_{10-40 \rm \, keV}$ $\approx$ $L_{2-10 \rm \, keV}$, the L$_{10-40  \rm \, keV}$, predicted from the above relations, lies only a factor of $\approx$ 2 above our measurement based upon spectral analysis of the X-ray data, which is a reasonable match given the many potential sources of uncertainty associated with corrections for large column densities. It should also be noted that the nominal WISE PSF is $\approx$ 6 arcsec (corresponding to a physical size of $\approx$ 5 kpc at the source redshift) and the observed WISE magnitude is likely to contain some contaminating emission from the host galaxy in addition to the AGN. The observed infrared (and hence, predicted X-ray powers) associated with the AGN alone will then be pushed down, and should agree even better with the directly measured intrinsic AGN luminosity

In conclusion, source ID~330 would not have been classified as a CT AGN by any means using \chandra\ or \xmm\ data alone or combining X-ray information with optical data since its spectral energy distribution is solely dominated by starlight. On the other hand, an X-ray to infrared correlation would have highlighted a possible discrepancy between the measured and predicted X-ray luminosity using \chandra\ or \xmm\ data only, however star-formation or the presence of an under-luminous AGN could have been used to explain it. Sensitive \nustar\ data at $>$10 keV were vital for the classification of this source as a CT AGN.


\begin{table*}
\centering
\caption{Spectral analysis results for source ID~330}
     \begin{threeparttable}

\begin{tabular}{l c c c c}
 \hline
 \hline
             & abs-pow	&   baseline	   &  MYTorus	     &   BNTORUS\\
\hline
$\Gamma$             &  1.1$\pm$0.1  &    2.1$\pm$0.2\tnote{a}  &   1.6$\pm$0.1\tnote{a} &    2.0$\pm$0.2\tnote{a}\\
 $N_{\rm H}$ (10$^{24}$ cm$^{-2}$)  & $<$2$\times 10^{-4}$	&    1.6$\pm$0.2  &   1.0$\pm$0.1     &    1.2$^{+0.3}_{-0.1}$\\
Cstat/dof         &  176.4/98	&   91.3/99	   &  106.0/101      &   92.9/101\\
$\Gamma_{scattered}$       &  \dotfill	&    2.1$\pm$0.2\tnote{a}  &   1.6$\pm$0.1\tnote{a}   &    2.0$\pm$0.2\tnote{a} \\
EW$_{Fe K\alpha}$ (eV)      &    \dotfill	&    237$^{+60}_{-65}$    &   316$\pm$10      &     \dotfill\\
Inclination Angle\tnote{b} (deg)         &    \dotfill	&60 	   &	85	     &    85  \\
Opening Angle (deg) &   \dotfill	&     \dotfill 	   &   $>$70\tnote{c} 	     &    77$^{+2}_{-9}$\tnote{c} \\
  $F_{3-8 \rm \, keV,  \nustar}$ (\cgs)     &  8.3$\times 10^{-14}$\tnote{d}  &   2.3$\times 10^{-14}$	   &  3.3$\times 10^{-14}$	     &   3.4$\times 10^{-14}$\\
  $F_{3-8 \rm \, keV, \chandra}$ (\cgs)      &  1.3$\times 10^{-14}$	&    2.6$\times 10^{-14}$	   &   2.3$\times 10^{-14}$       &    2.4$\times 10^{-14}$\\
  $F_{3-8 \rm \, keV, \xmm} $ (\cgs)     &  1.9$\times 10^{-14}$	&    3.3$\times 10^{-14}$	   &   3.0$\times 10^{-14}$       &    3.2$\times 10^{-14}$\\
  $L_{10-40 \rm \, keV, \nustar}$  (\lum) &  1.8$\times 10^{42}$	&    3.4$\times 10^{42}$	   &   4.7$\times 10^{42}$	  &    4.9$\times 10^{42}$\\
$L_{2-10 \rm \, keV, \nustar}$  (\lum) &  5.8$\times 10^{41}$	&    4.4$\times 10^{42}$	   &   2.8$\times 10^{42}$	  &    5.9$\times 10^{42}$\\
 \hline
\hline
\end{tabular}
\begin{tablenotes}
            \item[a] Gamma of the primary and scattered components are tied.
             \item[b] Fixed.
             \item[c] Measure obtained by additionally freeing the opening angle parameter during the fit.
            \item[d]{The higher \nustar\ flux is obtained by allowing a relative normalization constant factor to reach a value of 6.6. 
        Not allowing this degree of freedom in the model result in severe positive residuals especially in the \nustar\ band.}

        \end{tablenotes}
     \end{threeparttable}
     \label{ID330}

   \end{table*}

\section{Summary and Conclusion}

We have presented the 1.7 deg$^2$ \nustar\ COSMOS survey, the middle tier of the \nustar\ extragalactic ``wedding cake''. 
We employed $\sim$3 Ms of exposure time and 121 overlapping observations to obtain relatively uniform coverage, with an average, 
vignetting corrected exposure time of $\sim$100-120 ksec when summing FPMA and FPMB observations into a single mosaic (FPMA+B).

We tailored a multistep analysis procedure to optimally analyze the data in three bands, 3-24 keV, 3-8 keV and 8-24 keV, separately. 
First, extensive simulations were performed to test the detection and photometry strategy, to maximize completeness and reliability of detections, and to establish survey sensitivity. In particular, we used {\it SExtractor} to detect sources on probability maps, computed using data mosaics and background maps. 
Aperture photometry was performed to assess source significance. Deblending was applied to separate the contribution of multiple detections within 90$^{\prime\prime}$. 
Using the simulations, we estimated a positional uncertainty of $\sim$7$^{\prime\prime}$, which is energy independent, according to our findings. 
We determined probability thresholds of DET\_ML= 15.27, 14.99 and 16.17 in the 3-24, 3-8 and 8-24 keV bands, respectively, corresponding to 99\% reliability. The flux limit reached is 5.9 $\times$ 10$^{-14}$ \cgs\ in the 3-24 keV band, 2.9 $\times$ 10$^{-14}$ \cgs\ in the 3-8 keV band and 6.4 $\times$ 10$^{-14}$ \cgs\ in the 8-24 keV band at 20\% completeness. 
Second, the detection and photometry methods tested on simulations were applied to real data.
At the chosen DET\_ML thresholds, we detected 81, 61 and 32 sources in each band, for a total of 91 detections. The number of spurious sources in the full catalog is expected to be $\sim$2.

Thanks to the full coverage of the \nustar\ survey at lower energy by \chandra\ and \xmm\, we can associate a point-like, lower energy counterpart (including all the multiwavelength information available) to 87 sources. One source is matched to a \chandra\ and \xmm\ extended source, whose X-ray emission is consistent with thermal emission from hot extended gas in a galaxy cluster. Three sources remain unidentified at lower energies. Two are not detected in the 8-24 keV band, and only one (ID 245) is detected (but below threshold) in this band. Although variability and obscuration may explain the nature of these sources, their number is consistent with the expected number of spurious detections.

The detected sources span four decades in luminosity and cover a redshift range of $z$=0.04-2.5, extending to both faint luminosities and higher redshifts with respect to previous samples at energies  $>$ 10 keV. The sample consists of half unobscured AGN, classified either from optical spectroscopy or SED fitting, and half obscured AGN. 
In the X-rays, we used the hardness ratio to obtain an estimate on the observed fraction of candidate CT ($N_{\rm H} >10^{24}$ cm$^{-2}$) AGN of $\sim$13\% (with an upper value of 20\%) over the whole redshift and luminosity range covered by this survey, consistent with previous works in the hard X-rays (e.g. Krivonos et al. 2007, Tueller et al. 2008, Ajello et al. 2008, Burlon et al. 2011, Ajello et al. 2012, Fiore et al. 2012, Vasudevan et al. 2013). A more detailed analysis on the observed and intrinsic fraction of CT AGN and obscured sources in the COSMOS sample, including correction for absorption bias, will be presented by Zappacosta et al. (in prep.), while comparison with model predictions will be presented in Aird et al. (in prep.). 

According to their hardness ratio, two sources (ID~330 and ID~557) are classified as extremely obscured with $N_{\rm H} >$  10$^{24}$ cm$^{-2}$. 
While source ID~557 does not have enough counts for a reliable spectral analysis, source ID~330, a low X-ray luminosity spiral galaxy at $z$=0.044, is classified as heavily obscured and consistent with being a CT AGN. Without \nustar\ data, the source would not have been classified as such. 

For the first time, we present the the X-ray to optical flux ratio locus, originally defined in the soft band, in the 8-24 keV band using \nustar\ fluxes. About 10\% of the sources occupy a region above the locus defined with \chandra\ data in the 2-10 keV band, suggesting these are obscured AGN at high redshift with faint optical counterparts. Being at very low redshift, source ID~330 has the lowest X-ray to optical flux ratio, and can be associated to the class of sources named XBONGs, whose obscured nature has been previously suggested.

Given the sensitivity and volume of the COSMOS \nustar\ survey, the data are particularly valuable for probing the variety of sources contributing to the cosmic XRB at energies above $>$8 keV with fainter luminosities and higher redshifts than previous surveys. This survey complements that performed in the ECDFS, which provides a sample at a deeper flux limit, fundamental to constrain the number counts (Harrison et al. in prep.) and the resolved fraction of the XRB (Hickox et al. in prep.) at \nustar\ energies. 

The analysis performed in this paper is limited to the 24 keV energy, but we foresee extending the work to higher energy ($>$30 keV), performing both source detection to reveal bright sources with high energy spectra, and also stacking analysis of \chandra\ and \xmm\ sources. The sample presented here will be used together with the sources detected in the 4 tiers of the \nustar\ wedding cake (about 200 sources total) for the computation of the number counts, the X-ray luminosity function and the resolved fraction of the X-ray background at energies above 8 keV in Harrison et al. (in prep.) and Aird et al. (in prep.). In addition, Zappacosta et al. and Del Moro et al. (in prep.) will present the the joint \nustar, \xmm\ and \chandra\ spectral analysis deriving the X-ray spectral properties (spectral slope and column density) of both COSMOS and ECDFS bright and faint sources.

\section{Acknowledgement}
We thank the anonymous referee for the interesting comments and A. Goulding and M. Rose for useful discussions. 
This work made use of data from the \nustar\ mission, a project led by
the California Institute of Technology, managed by the Jet Propulsion
Laboratory, and funded by the National Aeronautics and Space
Administration. We thank the \nustar\ Operations, Software and
Calibration teams for support with the execution and analysis of these
observations. This research has made use of the \nustar\ Data Analysis
Software (NUSTARDAS) jointly developed by the ASI Science Data Center
(ASDC, Italy) and the California Institute of Technology (USA).  
We acknowledge support from the NASA grants 11-ADAP11-0218 and GO3-14150C (FC);  from the Science and
Technology Facilities Council ST/I001573/1 (ADM, DMA); NSF award AST 1008067 (DRB); NuSTAR grant 44A-1092750, NASA ADP grant NNX10AC99G, and the
V.M. Willaman Endowment (WNB, BL); CONICYT-Chile grants Basal-CATA PFB-06/2007 (FEB), FONDECYT 1141218 (FEB), and "EMBIGGEN" Anillo ACT1101 (FEB, ET);
the Ministry of Economy, Development, and Tourism's Millennium Science Initiative through grant IC120009, awarded to The Millennium Institute of Astrophysics, MAS (FEB);  the Center of Excellence in Astrophysics and Associated Technologies (PFB 06) and by the FONDECYT regular grant 1120061 (ET); financial support under ASI/INAF contract I/037/12/0 (LZ).

{}

\appendix

\section{Catalog Description}

The electronic version of the catalog will contain the properties as listed in Table 5.
All positions are Right Ascension and Declination in the J2000 coordinate system. 
The positional error is of 7$^{\prime\prime}$, as obtained from the simulation analysis and explained in Section \ref{distances}.
\begin{itemize}
\item[-] {\em Column 1:} \nustar\ source name, following the standard IAU
  convention with the prefix ``NuSTAR''.
\item[-] {\em Column 2:} Source number. Sources are listed in order of
  detection: first those detected in the 3-24 keV band,
  followed by those detected in the 3-8 keV band only and then by those detected in the 8-24 keV band only.
\item[-] {\em Column 3-4:} The X-ray coordinates of the source.
\item[-] {\em Column 5:} The 3-24 keV band deblended DET\_ML.
\item[-] {\em Column 6:} The 3-24 keV band exposure time at the position of the source.
\item[-] {\em Column 7:} The 3-24 keV band total counts in a 20$^{\prime\prime}$ radius aperture.
\item[-] {\em Column 8:} The 3-24 keV band deblended background counts in a 20$^{\prime\prime}$ radius aperture.
\item[-] {\em Column 9:} The 3-24 keV band not deblended  background counts in a 20$^{\prime\prime}$ radius aperture.
\item[-] {\em Column 10:} The 3-24 keV band net counts (deblended if detected or 3$\sigma$ upper limit) in a 20$^{\prime\prime}$ radius aperture.
\item[-] {\em Column 11:} The 3-24 keV band count error  computed using Gehrels statistic.
\item[-] {\em Column 12:} The 3-24 keV band count rate (90\% confidence upper limit if negative for undetected sources) in a 20$^{\prime\prime}$ radius aperture.
\item[-] {\em Column 13:}The  3-24 keV band aperture corrected flux (90\% confidence upper limit if negative for undetected sources).
\item[-] {\em Column 14:} The 3-24 keV band flux error (-99 for upper limits).
\item[-] {\em Column 15:} The 3-8  keV band deblended DET\_ML.
\item[-] {\em Column 16:} The 3-8  keV band exposure time at the position of the source.
\item[-] {\em Column 17:} The 3-8  keV band  total counts in a 20$^{\prime\prime}$ radius aperture.
\item[-] {\em Column 18:} The 3-8  keV band deblended background counts in a 20$^{\prime\prime}$ radius aperture.
\item[-] {\em Column 19:} The 3-8  keV band not deblended background counts  in a 20$^{\prime\prime}$ radius aperture.
\item[-] {\em Column 20:} The 3-8  keV band net aperture counts (deblended if detected or 3$\sigma$ upper limit) in a 20$^{\prime\prime}$ radius aperture.
\item[-] {\em Column 21:} The 3-8  keV band count error  computed using Gehrels statistic.
\item[-] {\em Column 22:} The 3-8  keV band count rate (90\% confidence upper limit if negative for undetected sources) in a 20$^{\prime\prime}$ radius aperture.
\item[-] {\em Column 23:}The  3-8  keV band  aperture corrected  flux (90\% confidence upper limit if negative for undetected sources).
\item[-] {\em Column 24:} The 3-8  keV band flux error (-99 for upper limits).
\item[-] {\em Column 25:} The 8-24  keV band deblended DET\_ML.
\item[-] {\em Column 26:} The 8-24  keV band exposure time at the position of the source.
\item[-] {\em Column 27:} The 8-24  keV band  total counts in a 20$^{\prime\prime}$ radius aperture.
\item[-] {\em Column 28:} The 8-24  keV band deblended background counts  in a 20$^{\prime\prime}$ radius aperture.
\item[-] {\em Column 29:} The 8-24  keV band not deblended aperture background counts in a 20$^{\prime\prime}$ radius aperture.
\item[-] {\em Column 30:} The 8-24  keV band net aperture counts (deblended if detected or 3$\sigma$ upper limit) in a 20$^{\prime\prime}$ radius aperture.
\item[-] {\em Column 31:} The 8-24  keV band count error  computed using Gehrels statistic.
\item[-] {\em Column 32:} The 8-24  keV band count rate (90\% confidence upper limit if negative for undetected sources) in a 20$^{\prime\prime}$ radius aperture.
\item[-] {\em Column 33:}The  8-24  keV band  aperture corrected  flux (90\% confidence upper limit if negative for undetected sources).
\item[-] {\em Column 34:} The 8-24  keV band flux error (-99 for upper limits).
\item[-] {\em Column 35:} Hardness ratio computed using BEHR and counts in a 20$^{\prime\prime}$ radius aperture.
\item[-] {\em Column 36:} HR  Lower Bound.
\item[-] {\em Column 37:} HR  Upper Bound.
\item[-] {\em Column 38:} Band ratio computed using BEHR and counts in a 20$^{\prime\prime}$ radius aperture.
\item[-] {\em Column 39:} Band ratio  lower bound.
\item[-] {\em Column 40:} Band ratio upper bound.
\item[-] {\em Column 41:} C-COSMOS identification number (see Elvis et al. 2009).
\item[-] {\em Column 42:} {\it XMM}-COSMOS identification number (see Brusa et al. 2010).
\item[-] {\em Column 43-44:} \chandra\ or \xmm\ X-ray coordinates of the associated source.
\item[-] {\em Column 45:} 0.5-2 keV flux from C-COSMOS and if not from {\it XMM}-COSMOS.
\item[-] {\em Column 46:} 2-10 keV flux from C-COSMOS and if not from {\it XMM}-COSMOS.
\item[-] {\em Column 47:} Spectroscopic redshift of the C-COSMOS or {\it XMM}-COSMOS counterpart.
\item[-] {\em Column 48:} Spectroscopic source type classification (1= broad line AGN, 2 = narrow emission line)
\item[-] {\em Column 49:}  Photometric redshift of the C-COSMOS or {\it XMM}-COSMOS counterpart.
\item[-] {\em Column 50:}  Photometric source type classification (1= unobscured AGN, 2 = obscured AGN; see Civano et al. 2012 for details).
\item[-] {\em Column 51:} \nustar\ to \chandra\ or \xmm\ position separation.
\item[-] {\em Column 52:} Luminosity Distance in Mpc using the spectroscopic redshift.
\item[-] {\em Column 53:} Luminosity Distance in Mpc using the photometric redshift.
\item[-] {\em Column 54:} 3-24 keV luminosity (upper limit if negative number).
\item[-] {\em Column 55:} 3-8 keV luminosity (upper limit if negative number).
\item[-] {\em Column 56:} 8-24 keV luminosity (upper limit if negative number).
\item[-] {\em Column 57:} Flag for sources with multiple low energy counterparts (0=false, 1=true).

\end{itemize}

\end{document}